\documentclass[11pt]{article}

\pdfoutput=1
\usepackage{yfonts}
\usepackage{color}
\usepackage{mhchem}
\usepackage{xcolor}
\usepackage{cite}
\usepackage{sectsty}
\usepackage{bm}
\usepackage{array}

\usepackage{environ}
\NewEnviron{myalign}{%
\begin{align}
\scalebox{1.15}{$\BODY$}
\end{align}
}
\NewEnviron{smalign}{%
\begin{align}
\scalebox{0.85}{$\BODY$}
\end{align}
}
\usepackage{yfonts}
\usepackage{color}
\usepackage{mhchem}
\usepackage{xcolor}
\usepackage{mathrsfs}
\usepackage[mathscr]{eucal}
\usepackage{cite}
\usepackage{hyperref}
\hypersetup{colorlinks=true,linkcolor=red,anchorcolor=black,citecolor=green}
\usepackage[toc,page]{appendix}
\usepackage{amsfonts}
\usepackage{bbold}
\usepackage{textcomp}
\usepackage[DIV13]{typearea}
\usepackage{amsmath, amsthm, amssymb, mathtools,empheq,latexsym,dsfont}
\usepackage{bbm}
\usepackage{slashed, simplewick}
\usepackage[utf8]{inputenc}
\usepackage{graphicx,placeins}
\usepackage{makeidx}
\usepackage[font=small,labelfont=bf]{caption}
\usepackage{nicefrac}
\usepackage{subfigure}
\usepackage{array, bigdelim,multirow,multicol}
\usepackage[integrals]{wasysym}
\usepackage{fancybox}
\usepackage{bm}
\usepackage{float}
\usepackage{rotating}
\usepackage{colortbl}
\usepackage{booktabs}
\usepackage{longtable}
\usepackage{lscape}
\usepackage[top=2cm,textwidth=16.6cm,textheight=23.75cm]{geometry}
\usepackage{doi}
\usepackage{amsmath}
\allowdisplaybreaks[4]
\newcommand{\ignore}[1]{}
\graphicspath{{immagini/}}

\begin{document}

\title{
\begin{flushright}
\begin{minipage}{0.2\linewidth}
\normalsize
\end{minipage}
\end{flushright}
{\Large \bf Modular Symmetry with Weighton \\[2mm]}
\date{}
\author{
Gui-Jun Ding$^{a}$\footnote{E-mail: {\tt
dinggj@ustc.edu.cn}},  \
Stephen F. King$^{b}$\footnote{E-mail: {\tt king@soton.ac.uk}}, \
Jun-Nan Lu$^{a}$\footnote{E-mail: {\tt
hitman@mail.ustc.edu.cn}},  \
Ming-Hua Weng$^{c}$\footnote{E-mail: {\tt
mhweng@mju.edu.cn}} \
\\*[20pt]
\centerline{
\begin{minipage}{\linewidth}
\begin{center}
$^a${\it \small Department of Modern Physics,  and Anhui Center for fundamental sciences in theoretical physics,\\
University of Science and Technology of China, Hefei, Anhui 230026, China}\\[2mm]
$^b${\it \small Physics and Astronomy, University of Southampton, Southampton, SO17 1BJ, U.K.}\\[2mm]
$^c${\it \small College of Physics and Electronic Information Engineering, Minjiang University,\\ Fuzhou 350108, China}
\end{center}
\end{minipage}}
\\[10mm]}}

\maketitle
\thispagestyle{empty}

\begin{abstract}
We systematically develop the weighton mechanism for natural quark and charged lepton mass hierarchies in the framework of modular symmetry with a single modulus field $\tau$. The weighton $\phi$ is defined as a complete singlet with unit modular weight, leading to fermion mass suppression by powers of $\tilde{\phi}$, which is the vacuum expectation value of the field scaled by a flavour cut-off. Further mass and mixing angle suppression comes from powers of the small parameter, $q\equiv e^{i2\pi \tau}$. Assuming some fields transform as triplets under the finite modular symmetry, with general assignments for the other fields, we perform a complete analysis for the levels $N=3, 4, 5$, expressing fermion masses and mixings in terms of powers of the small parameters $\tilde{\phi}$ and $q$. We present two examples in detail, based on the modular group $T'$, close to the CP boundary of $\tau$, which can address both fermion mass and mixing hierarchies using a weighton field.

\end{abstract}

\clearpage

\section{\label{sec:introduction}Introduction}

Most of free parameters in the Standard Model (SM) appear in the flavor sector, they are the masses, mixing angles and CP violation phases of quarks and leptons. The observed masses of quarks and leptons as well as their mixing reveal a peculiar pattern. It is well-known that there are large mass hierarchies among the three generations of up quarks, down quarks, and charged leptons, the mass spectra have a span of at least 5 orders of magnitude from the electron mass about 0.511 MeV to the heavy top quark mass approximately 173 GeV. The quark flavor mixing is described by the Cabibbo-Kobayashi-Maskawa (CKM) matrix which is approximately an identity matrix,
and the three quark mixing angles are hierarchically suppressed with $\theta^q_{12}\sim\lambda\gg\theta^q_{23}\sim\lambda^2\gg\theta^q_{13}\sim\lambda^3$, where $\lambda\simeq0.225$ denotes the Cabibbo angle~\cite{ParticleDataGroup:2024cfk}. In comparison with the quarks and charged leptons, neutrino oscillation shows that neutrinos have tiny masses which are at least six orders of magnitude smaller than the electron mass. Unlike the small and hierarchical mixing between quarks, neutrino mixing exhibit large mixing angles.

In SM, the quark and charged lepton masses arise from three independent Yukawa couplings $Y_{u}$, $Y_d$ and $Y_{l}$ which are $3\times3$ matrices in flavor space, and the value of each entry of $Y_{u}$, $Y_d$, $Y_{l}$ is not subject to any constraint from the SM gauge symmetry. However, the hierarchical quark masses requires the singular values of $Y_{u}$, $Y_d$, $Y_{l}$ have a similar hierarchical structure. Moreover, the observed CKM matrix implies an approximate alignment between $Y_{u}$ and $Y_d$. The particular structure of $Y_{u}$ and $Y_d$ clearly calls for an explanation. Neutrinos are massless in the minimal renormalizable SM, consequently SM must be extended to explain the non-zero neutrino masses. The interactions responsible for the neutrino masses should have certain structure to accommodate the large lepton mixing.

There have been many attempts to explain the masses and flavor mixing patterns of quarks and leptons~\cite{King:2017guk,Petcov:2017ggy,Xing:2020ijf,Feruglio:2019ybq,Ding:2024ozt}. Modular symmetry is a very appealing approach to address the flavor structure of SM~\cite{Feruglio:2017spp}. It provides a natural origin of the discrete flavor symmetry, and the modular invariance enforces the Yukawa couplings to be holomorphic modular forms in the context of supersymmetry. The vacuum expectation value (VEV) of the complex modulus $\tau$ is the unique source breaking modular symmetry, so that the flavor symmetry breaking sector is minimized. The various aspects of modular flavor symmetry have been comprehensively reviewed in~\cite{Kobayashi:2023zzc,Ding:2023htn}. The modular symmetry models are quite economic and predictive, it is remarkable that all the lepton masses, mixing angles and CP violation phases can be explained by only four real parameters besides the modulus $\tau$ in the most predictive modular invariant model~\cite{Ding:2022nzn,Ding:2023ydy}. The quark sector can be incorporated at the expense of introducing eight more real parameters~\cite{Ding:2022nzn,Ding:2023ydy}, and the modulus $\tau$ is the link connecting quark and lepton sectors so that the flavor observables of quarks and leptons could be strongly correlated with each other~\cite{Ding:2023ydy,Ding:2024pix}. The modulus $\tau$ could also play an important rule in Cosmology, it could play the role of inflaton~\cite{Conlon:2005jm,Ben-Dayan:2008fhy,Kobayashi:2016mzg,Abe:2023ylh,Ding:2024neh,King:2024ssx,Casas:2024jbw,Kallosh:2024ymt}, and it could also be responsible for the reheating of the Universe~\cite{Ding:2024euc}.

In the phenomenologically viable flavor models based on modular symmetry, one usually needs to fine-tune the values of some couplings in the models in order to accommodate the mass hierarchies of the charged leptons and quarks. The hierarchy of charged lepton and quark masses can follow from the properties of the modular forms present in the fermion mass matrices, if the VEV of $\tau$ is close to the modular symmetry fixed points $\tau_0=e^{2\pi i/3}$ or $\tau_0=i\infty$~\cite{Okada:2020ukr,Novichkov:2021evw,Petcov:2022fjf,Kikuchi:2023cap,Kikuchi:2023jap,Abe:2023qmr,Petcov:2023vws,deMedeirosVarzielas:2023crv}, but not the third modular fixed point in the fundamental domain $\tau_0=i$. In the vicinity of a fixed point, the theory still enjoys a residual symmetry\footnote{The residual symmetry is $Z^S_4$, $Z^{ST}_3\times Z^{R}_2$ and $Z^{T}_N\times Z^R_2$ with $R=S^2$ in the vicinity of $\tau_0=i$, $\tau_0=e^{2\pi i/3}$ and $\tau_0=i\infty$ respectively, and it is spontaneously broken by $\tau-\tau_0$.}. As a result, the fermion mass matrices and modular forms are strongly constrained by this residual symmetry if $\tau$ is close to $\tau_0$, and each entry of the fermion mass matrices is certain power of the small departure $|\tau-\tau_0|$ in proper basis~\cite{Novichkov:2021evw}. Hence, hierarchical fermion mass matrices arise solely due to the proximity of the modulus $\tau$ to the residual symmetry point $\tau_0$. However, for the modulus near $\tau_0=e^{2\pi i/3}$ or $\tau_0=i\infty$, the lepton mixing angles require the left-handed leptons to be singlets of the finite modular group~\cite{Novichkov:2021evw}, and consequently more parameters are introduced in the resulting models~\cite{Okada:2020ukr,Novichkov:2021evw,Petcov:2022fjf,Kikuchi:2023cap,Kikuchi:2023jap,Abe:2023qmr,Petcov:2023vws,deMedeirosVarzielas:2023crv}.
Conversely, for the modulus near $\tau_0=i$,
left-handed leptons may be triplets of the finite modular group, while allowing
lepton mixing angles and neutrino mass squared differences to be naturally accommodated~\cite{Feruglio:2022koo,Feruglio:2023mii},
but in this case it is not possible to reproduce the charged lepton mass hierarchies without fine-tuning the dimensionless couplings~\footnote{Interestingly, for the modulus near $\tau_0=i$,
it is found that all physical quantities exhibit a universal scaling with the distance $|\tau-\tau_0|$, and there is no dependence on the level of the construction, the weights and representation assignments of matter fields, and even the form of the kinetic terms~\cite{Feruglio:2021dte,Feruglio:2022koo,Feruglio:2023mii}. The same universal behavior is found in Siegel modular invariant theories with multiple moduli near the fixed points~\cite{Ding:2024xhz}.}.

Fermion mass hierarchies may also be addressed by the weighton mechanism~\cite{King:2020qaj}, in which the modular weights of fermion fields play the role of Froggatt-Nielsen (FN) charges~\cite{Froggatt:1978nt}, and an additional weighton field, which is both an SM singlet and a finite modular symmetry singlet with non-zero modular weight, plays the role of FN flavons (although singlet weighton VEVs do not break the symmetry, unlike FN flavons~\footnote{A suppression mechanism involving powers of finite modular symmetry {\it triplets} was previously considered to account for charged lepton mass hierarchies in~\cite{Criado:2019tzk}, but here we shall be interested in weightons, which by definition are complete {\it singlets} which carry modular weights.}). In the realistic weighton models of
quarks and leptons discussed in~\cite{King:2020qaj},
both numerical and analytic results were presented, using an expansion in both the weighton suppression and the $q$-expansion, as discussed below, with results presented for a value of $\tau$ close to imaginary axis, where CP is conserved, and two $A_4$ benchmark examples were discussed.

In the present paper we systematically develop the weighton mechanism proposed in~\cite{King:2020qaj}, classifying and providing analytic results for large classes of models at levels $N=3,4,5$ where at least some fields (for example the lepton doublets) transform as triplets of the finite modular symmetry, and where the quark and lepton mixing and CP violation can be achieved together with natural quark and charged lepton mass hierarchies. As in~\cite{King:2020qaj},
there are three key ingredients essential to the success of the approach. The first is the intrinsic hierarchical nature of modular forms, as seen in their $q$-expansion, which we systematically exploit to estimate the orders of magnitude of the Yukawa couplings, once the representation assignments of matter fields are specified.
The second is the use of the weighton mechanism to supplement the modular symmetry, which seems inevitable in order to achieve fully natural hierarchical patterns of fermion mass matrices.
The third is the identification of the CP boundary as the natural region to work, since large CP violation in the vicinity of this boundary can then be obtained without disturbing the successful mass and mixing predictions on the boundary.
Concerning the first ingredient,
it is known that the modular forms can be expressed as a convergent infinite series of $q\equiv e^{2\pi i\tau}$, this is the so-called $q$-expansion. In the fundamental domain, we have $|q|\leq e^{-\sqrt{3}\pi}\simeq0.0043\ll1$ which is a good expansion parameter. By considering the modular invariance under the $T$ transformation, we show how it is possible to estimate the order of magnitude of the Yukawa couplings parameterized by powers of $q$, once the representation assignments of matter fields are specified. However, as discussed, this general property of modular forms is not enough by itself to naturally account for the fermion mass hierarchies.
When the second ingredient, the weighton field $\phi$, is included, each Yukawa coupling can be expressed as powers of $q$ and $\tilde{\phi}\equiv\langle\phi\rangle/M_{fl}$ at leading order, where $M_{fl}$ is the cut-off flavour scale, leading to natural hierarchical patterns of quark and charged lepton mass matrices, together with successful mixing. When addressing the problem of CP violation, we introduce the third ingredient, which is to work either near the imaginary axis or the boundary of the fundamental domain, which are the places where CP is  conserved. We refer to these two lines as the CP boundary (see figure~\ref{fig:gCP-fixed-points}). On the CP boundary, the theory enjoys a residual CP symmetry, and if the complex modulus takes a value in the vicinity of the CP fixed points, this residual CP symmetry strongly constrains the phases the Yukawa couplings, and even a small deviation from the CP fixed points can lead to large CP violation. Thus working near the CP boundary conveniently decouples the problem of fermion mass and mixing from that of CP violation.

The layout of the remainder of this paper is as follows. We give a brief introduction to modular symmetry in section~\ref{sec:modular-CP-residual-symmetries}. The general constraints on the fermion mass matrix by modular symmetry and weightons are analyzed in section~\ref{sec:modular-mass-matrix}. By considering the invariance under the modular $T$ transformation, we can estimate the order of the magnitude of each Yukawa coupling in both scenarios with and without weighton. We perform a comprehensive analysis of the possible hierarchical patterns of fermion masses and rotation angles for $N=3, 4, 5$ in section~\ref{sec:hierarchical-patterns-Nleq5}, and the results are all tabulated in Appendix~\ref{app:patterns-masses}. Two example models based on the $T'$ modular symmetry are presented in section~\ref{sec:example-models}, where the VEV of $\tau$ is near the imaginary axis and the left vertical boundary of the fundamental domain respectively, and the mass and mixing hierarchies of both quarks and leptons can be addressed, including detailed numerical fits. Finally we summarize the results and conclude in section~\ref{sec:conclusion}. The finite modular group $\Gamma'_3\cong T'$ and the modular forms of level 3 are given in the Appendix~\ref{app:Tp-group}. We collect the possible patterns of fermion masses and rotation angles for the levels $N=3, 4, 5$ in Appendix~\ref{app:patterns-masses}, where at least either the electroweak left-handed fields or the charge conjugation of right-handed fields transform as a triplet of the finite modular group $\Gamma'_N$. We discuss the general constraints on the Yukawa  by the generalized CP (gCP) symmetry and how one can naturally generate the CP violation phases in the vicinity of residual CP symmetric points in section~\ref{sec:Yukawa-CP-fixed-points}.

\section{\label{sec:modular-CP-residual-symmetries}Modular symmetry and residual symmetry}

The modular symmetry is ubiquitous in compactifications of String theory, and it is geometrical symmetry of the extra compact space. In the simple toroidal compactification, the two-dimensional torus $T^2$ is described as the quotient $T^{2}=\mathbb{C}/\Lambda_{\omega_1, \omega_2}$, where $\mathbb{C}$ refers to the whole complex plane $\mathbb{C}$ and $\Lambda_{\omega_1, \omega_2}=\left\{m\omega_1+n\omega_2, m,n\in\mathbb{Z}\right\}$ denotes a two-dimensional lattice with the basis vectors $\omega_1$ and $\omega_2$. The lattice is left invariant under a change of the basis vectors
only and if only
\begin{equation}
\begin{pmatrix}
\omega_1\\
\omega_2
\end{pmatrix}\rightarrow\begin{pmatrix}
\omega'_1\\
\omega'_2
\end{pmatrix}=\begin{pmatrix}
a  ~&~  b \\
c  ~&~ d
\end{pmatrix}\begin{pmatrix}
\omega_1\\
\omega_2
\end{pmatrix}\,,
\end{equation}
with $a$, $b$, $c$, $d$ integers and $ad-bc=1$. The geometry of a torus is characterized by the complex modulus $\tau=\omega_1/\omega_2$ up to rotation and scale transformations, without loss of generality we can limit $\tau$ to the upper half complex plane with $\text{Im}(\tau)>0$. The two tori related by modular transformations would be identical, i.e.
\begin{equation}
\label{eq:modular-trans}\tau\xrightarrow{\gamma}\tau'=\frac{\omega'_1}{\omega'_2}=\frac{a\tau+b}{c\tau+d}\equiv\gamma\tau,~~~\gamma=\begin{pmatrix}
a  ~&~  b \\
c  ~&~ d
\end{pmatrix},~~~\text{Im}(\tau)>0\,.
\end{equation}
We see that each modular transformation corresponds to a two-dimensional integer matrix $\gamma$ of unit determinant. Thus the modular group is isomorphic to the special linear group $SL(2, \mathbb{Z})\equiv \Gamma$. Moreover, from Eq.~\eqref{eq:modular-trans} we see that $\gamma\in\Gamma$ and $-\gamma\in\Gamma$ give the same transformation of $\tau$. The modular group $\Gamma$ has infinite elements, it can be generated by two generators $S$ and $T$ with
\begin{equation}
S=\begin{pmatrix}
0 ~& 1 \\
-1 ~& 0
\end{pmatrix},~~~T=\begin{pmatrix}
1 ~& 1 \\
0 ~& 1
\end{pmatrix}\,,
\end{equation}
which satisfy the relations $S^4=(ST)^3=\mathbb{1}$ and $S^2T=TS^2$. The action of the generators $S$ and $T$ on the complex modulus is
\begin{equation}
\tau\xrightarrow{S}-\frac{1}{\tau},~~~\tau\xrightarrow{T}\tau+1\,.
\end{equation}
Employing modular transformations, one can any point in the upper half complex plane to the fundamental domain $\mathcal{D}$ defined by
\begin{eqnarray}
\label{eq:fundamental-domain}\mathcal{D}=\left\{\tau\,\Big|\,\text{Im}(\tau)>0, |\text{Re}(\tau)|\leq \frac{1}{2}, |\tau|\geq1\right\}\,.
\end{eqnarray}
Notice that the right half of the boundary of $\mathcal{D}$ is related to the left half by the $T$ transformation, but no two points in the interior of $\mathcal{D}$ are related under the modular transformation. It is sufficient to focus on the values of the complex modulus $\tau$ in the fundamental domain $\mathcal{D}$.

We work in the framework of global supersymmetry in this work, the matter fields such as leptons and quarks are denoted by chiral superfields. Under the action of modular group, the matter fields $\Phi_{I}$ transform as~\cite{Feruglio:2017spp,Lauer:1989ax,Ferrara:1989qb,Ferrara:1989bc}
\begin{equation}
\Phi_{I}\xrightarrow{\gamma}(c\tau+d)^{-k_I}\rho_{I}(\gamma)\Phi_I\,,
\end{equation}
where $k_I\in\mathbb{Z}$ is the modular weight, and $\rho_{I}$ is a unitary representation of the homogeneous finite modular group $\Gamma'_N=\Gamma/\Gamma(N)\cong SL(2, \mathbb{Z}_N)$ or the inhomogeneous finite modular group $\Gamma_N=\Gamma/\pm\Gamma(N)$. Here $\Gamma(N)$ is the principal congruence subgroup of level $N$. The superpotential $\mathcal{W}$ responsible for the fermion mass can be expanded in powers of matter superfield $\Phi_I$ as follow,
\begin{equation}
\mathcal{W}(\tau, \Phi_I)=\sum_{n}\left(Y_{I_1\ldots I_n}(\tau)\Phi_{I_1}\ldots\Phi_{I_n}\right)_{\mathbf{1}}\,,
\end{equation}
where one should sum over all possible field combinations and independent singlet contractions. Modular invariance of the superpotential requires that the function $Y_{I_1\ldots I_n}(\tau)$ should be modular forms of level $N$ of weight $k_Y=k_{I_1}+\ldots+k_{I_n}$, and its modular transformation is
\begin{equation}
Y_{I_1\ldots I_n}(\tau)\xrightarrow{\gamma}Y_{I_1\ldots I_n}(\gamma\tau)=(c\tau+d)^{k_Y}\rho_{Y}(\gamma)Y_{I_1\ldots I_n}(\tau)\,,
\end{equation}
where $\rho_{Y}$ is a unitary representation of $\Gamma'_N$ or $\Gamma_N$ and it satisfies $\rho_Y\otimes\rho_{I_1}\otimes\ldots\otimes\rho_{I_n}\supset\mathbf{1}$. Hence modular symmetry constrains the Yukawa couplings to be modular forms of level $N$, which span a linear space of finite dimension.

Assuming global supersymmetry, the K\"ahler potential should be modular invariant up to a K\"ahler transformation. Although the most general form of K\"ahler potential can involve modular forms and many terms are compatible with modular invariance in bottom-up approach~\cite{Chen:2019ewa}, usually the following the minimal K\"ahler potential is adopted~\cite{Feruglio:2017spp},
\begin{equation}
\mathcal{K}(\tau, \bar{\tau}, \phi_I, \bar{\phi}_I)=-\Lambda^2_0\log(\text{Im}\tau)+\sum_I\frac{\bar{\Phi}_I\Phi_I}{(\text{Im}\tau)^{k_I}}\,,
\end{equation}
where $\Lambda_0$ is a parameter with mass dimension one. After the modulus $\tau$ gets a VEV, the above K\"ahler potential gives rise to the following kinetic terms for the scalar components $\phi_I$ of the supermultiplets $\Phi_I$ and the modulus field $\tau$,
\begin{equation}
\mathcal{L}_K=\frac{\Lambda^2_0}{4\langle\text{Im}\tau\rangle^2}\partial_{\mu}\bar{\tau}\partial^{\mu}\tau+\sum_I\frac{\partial_{\mu}\bar{\phi}_I\partial^{\mu}\phi_I}{\langle\text{Im}\tau\rangle^{k_I}}\,.
\end{equation}
Consequently one needs to rescale the superfields $\Phi_I\rightarrow\langle\text{Im}\tau\rangle^{k_I/2}\Phi_I$ to make the kinetic terms normalized canonically.

In modular invariant models, the VEV of the complex modulus $\langle\tau\rangle$ is the unique source of flavour symmetry breaking. It is known that there is no value of $\langle\tau\rangle$ which is invariant under the whole modular transformation group. However, if $\langle\tau\rangle$ is at some fixed point $\langle\tau\rangle=\tau_f$, the modular symmetry $\Gamma$ would be broken down to a residual subgroup $\left\{\gamma^n_f: n\in\mathbb{Z}\right\}$ generated by the stabilizer $\gamma_f$ satisfying $\tau_f=\gamma_f\tau_f$. In the fundamental domain $\mathcal{D}$, there are only three independent modular fixed points: $\langle\tau\rangle=\tau_S=i$, $\langle\tau\rangle=\tau_T=i\infty$ and $\langle\tau\rangle=\tau_{ST}=(-1+i\sqrt{3})/2$, and the corresponding
residual modular symmetries are $Z^{S}_2\equiv\left\{1, S\right\}$, $Z^{T}\equiv\left\{1, T, \ldots\right\}$ and $Z^{ST}_3\equiv\left\{1, ST, (ST)^2\right\}$ respectively~\cite{Novichkov:2018yse,Ding:2019gof}. Note that $R\tau=\tau$ for any value of $\tau$, consequently $Z^R_2$ is always preserved.
If the modulus VEV $\langle\tau\rangle$ is exactly at the fixed point $\tau_f$, the charged lepton and neutrino mass matrices would enjoy the same residual flavor symmetry so that the lepton mixing matrix would be a unit matrix up to row and column permutations or be block-diagonal~\cite{Novichkov:2018yse,Ding:2019gof}. Hence $\langle\tau\rangle$ should depart from these modular symmetry fixed points in order to be phenomenologically viable. If $\langle\tau\rangle$ is in the vicinity of $\tau_S=i$, it was found that the large lepton mixing angles and the splitting between the solar and atmospheric neutrino mass squared differences can be naturally accommodated~\cite{Feruglio:2021dte,Feruglio:2022koo,Feruglio:2023mii,Ding:2024xhz}, and the theoretical predictions are independent from details of the models such as the modular weights and representation assignments of matter field under the finite modular groups. On the other hand, small departure from the other two fixed points $\tau_{ST}=(-1+i\sqrt{3})/2$ and $\tau_T=i\infty$ can naturally generate the observed mass hierarchies of charged leptons~\cite{Okada:2020ukr,Novichkov:2021evw,Petcov:2022fjf,Kikuchi:2023cap,Kikuchi:2023jap,Abe:2023qmr,Petcov:2023vws,deMedeirosVarzielas:2023crv}.
However, in all cases, it is difficult to reconcile the charged lepton mass hierarchies and large lepton mixing angles solely from the small deviation of $\langle\tau\rangle$ from the modular fixed point, if the three generations of left-handed lepton fields transform as a triplet of the finite modular group.

\section{\label{sec:modular-mass-matrix}The modular invariant fermion mass matrix }

The modular invariance imposes strong constraints on the Yukawa couplings. In this section we consider the charged lepton and neutrino mass terms for illustration. The methodology and conclusion are quite general, and they are applicable to the up-type and down-type quark mass terms. We assume that neutrinos are Majorana particles in the following. The modular invaraint superpotential for the charged lepton and neutrino masses in the framework of Minimal Supersymmetric Standard Model (MSSM) can be generally written as
\begin{equation}
\label{eq:Lag_mass}
\mathcal{W}=-E^{c}_i\mathcal{Y}^{e}_{ij}(\tau)L_jH_d-\frac{1}{2\Lambda}L_iL_j\mathcal{Y}^{\nu}_{ij}(\tau)H_uH_u\,,
\end{equation}
where the fields $L_i$ and $E^c_i=e^{c}, \mu^{c}, \tau^{c}$ stand for the left-handed lepton doublets and the right-handed charged leptons respectively, $H_u$ and $H_d$ are MSSM Higgs doublets, and $\mathcal{Y}^{e}_{ij}(\tau)$ and $\mathcal{Y}^{\nu}_{ij}(\tau)$ are modular forms. Under a generic modular transformation $\gamma\in\Gamma$, the lepton fields $L=\left(L_1, L_2, L_3\right)$, $E^c=\left(e^c, \mu^c, \tau^c\right)^{T}$ and the Higgs fields $H_{u,d}$  transform as
\begin{eqnarray}
\nonumber &L_i\stackrel{\gamma}{\mapsto}(c\tau+d)^{-k_{L_i}}\left(\rho_{L}(\gamma)\right)_{ij}L_j,~\quad~
&E^c_i\stackrel{\gamma}{\mapsto}(c\tau+d)^{-k_{E^c_i}}\left(\rho_{E^c}(\gamma)\right)_{ij}E^c_j\,, \,\\
& H_u\stackrel{\gamma}{\mapsto}(c\tau+d)^{-k_u}\rho_{u}(\gamma)H_u,~\quad~
&H_d\stackrel{\gamma}{\mapsto}(c\tau+d)^{-k_d}\rho_{d}(\gamma) H_d\,,
\end{eqnarray}
where $-k_L$, $-k_{E^c}$, $-k_u$ and $-k_d$ are the modular weights of the fields $L$, $E^c$, $H_u$ and $H_d$ respectively. Modular invariance requires that the summation of the modular weights of each term in Eq.~\eqref{eq:Lag_mass} should be vanishing and the Yukawa matrices $\mathcal{Y}^e\left(\tau\right)$ and $\mathcal{Y}^{\nu}\left(\tau\right)$ should transform under the action of modular symmetry as
\begin{eqnarray}
\nonumber&&\mathcal{Y}^e_{ij}\left(\tau\right)\stackrel{\gamma}{\mapsto}\mathcal{Y}^e_{ij}\left(\gamma\tau\right)=(c\tau+d)^{k_{E^c_i}+k_{L_j}+k_d}\rho^{*}_{d}(\gamma)\left[\rho^*_{E^c}(\gamma)\mathcal{Y}^e\left(\tau\right)\rho^{\dagger}_L(\gamma)\right]_{ij}\,,\,\\
\label{eq:me-mnu-gamma}&&\mathcal{Y}^\nu_{ij}\left(\tau\right)\stackrel{\gamma}{\mapsto}\mathcal{Y}^\nu_{ij}\left(\gamma\tau\right)=(c\tau+d)^{k_{L_i}+k_{L_j}+2k_u}\rho^{2*}_{u}(\gamma)\left[\rho^*_{L}(\gamma)\mathcal{Y}^\nu\left(\tau\right)\rho^{\dagger}_L(\gamma)\right]_{ij}\,.
\end{eqnarray}

\subsection{\label{qexpan} Formula for magnitudes of Yukawa couplings from the $T$ transformation}

The modular group has two generators $S$ and $T$, and the modular transformation $T$ has some special property. The modular transformation of the Yukawa couplings is given in Eq.~\eqref{eq:me-mnu-gamma}, taking $\gamma=T$, we have
\begin{eqnarray}
\nonumber\mathcal{Y}^e\left(\tau+1\right)&=&\rho^{*}_{d}(T)\rho^*_{E^c}(T)\mathcal{Y}^e\left(\tau\right)\rho^{\dagger}_L(T)\,,\,\\
\mathcal{Y}^\nu\left(\tau+1\right)&=&\rho^{2*}_{u}(T)\rho^*_{L}(T)\mathcal{Y}^\nu\left(\tau\right)\rho^{\dagger}_L(T)\,.
\end{eqnarray}
It is convenient to work in the basis where the representation matrix of the modular generator $T$ is diagonal basis. Since $T^{N}=1$ in the finite modular group, each diagonal entry of $\rho_{E^c}(T)$ and $\rho_L(T)$ must be some power of $\zeta\equiv e^{2\pi i/N}$ which is the $N$th root of unit. Hence $T$ transformation of the element $\mathcal{Y}^e_{ij}$ is
\begin{equation}
\mathcal{Y}^e_{ij}(\tau+1)
=\zeta^{-k_{ij}}\mathcal{Y}^e_{ij}(\tau)\,,
\label{eq:Yukawa-T-trans}
\end{equation}
where we have denoted $\rho_{d}(T)\rho_{E^{c}_i}(T)\rho_{L_{j}}(T)=\zeta^{k_{ij}}$ and  $k_{ij}$ are some integer between 0 and $N-1$. From Eq.~\eqref{eq:Yukawa-T-trans} and the identity $\zeta^N=1$, we see that $\mathcal{Y}^e_{ij}(\tau)$ is a periodic function with a period of $N$, i.e.
\begin{eqnarray}
\mathcal{Y}^e_{ij}(\tau+N)=\mathcal{Y}^e_{ij}(T^N\tau)=\mathcal{Y}^e_{ij}(\tau)\,.
\end{eqnarray}
Notice that the Yukawa coupling $\mathcal{Y}^e_{ij}(\tau)$ are modular forms of level $N$ in modular flavor symmetry, and this property of periodicity follows from the definition of modular forms. Hence the Fourier expansion of $\mathcal{Y}^e_{ij}(\tau)$ can be written as
\begin{equation}
\mathcal{Y}^e_{ij}(\tau)=\sum_{n=0}^{\infty}C_{n}^{ij}q^{n/N},~~~q\equiv e^{2\pi i\tau}\,,
\end{equation}
where $C_{n}^{ij}$ are some constants independent of $\tau$ since Yukawa couplings are modular forms which are holomorphic functions of $\tau$. It is straightforward to show that $q^{1/N}$ transforms under $T$ as follows
\begin{equation}
q^{1/N}\stackrel{T}{\longrightarrow}
e^{2\pi i(\tau+1)/N}=e^{2\pi i/N}q^{1/N}=\zeta q^{1/N},~~q\stackrel{T}{\longrightarrow}q\,.\label{eq:q-1/N-mF1}
\end{equation}
Thus the modular invariance under the $T$ transformation in Eq.~\eqref{eq:Yukawa-T-trans} implies
\begin{equation}
C_{n}^{ij}=0,~~n\neq -k_{ij}\, (\text{mod}~N)\,.
\end{equation}
Hence the Yukawa couplings $\mathcal{Y}^e_{ij}(\tau)$ are of the following form~\footnote{As explained in section~\ref{sec:modular-CP-residual-symmetries}, the matter superfields have to be rescaled to canonically normalize their kinetic terms, and these rescalings can be absorbed into the couplings $C_{N-k_{ij}}^{ij}$, $C_{2N-k_{ij}}^{ij}$, $C_{3N-k_{ij}}^{ij}$ etc.}
\begin{eqnarray}
\nonumber\mathcal{Y}^e_{ij}(\tau)&=&C_{N-k_{ij}}^{ij}q^{(N-k_{ij})/N}+C_{2N-k_{ij}}^{ij}q^{(2N-k_{ij})/N}+C_{3N-k_{ij}}^{ij}q^{(3N-k_{ij})/N}+\ldots\,\\
\label{Yeij-expansion}&=&q^{(N-k_{ij})/N}\left[C_{N-k_{ij}}^{ij}+C_{2N-k_{ij}}^{ij}q+C_{3N-k_{ij}}^{ij}q^2+\ldots\right]\,. \end{eqnarray}
It follows that the magnitude of the Yukawa coupling $\mathcal{Y}^e_{ij}$ is determined by the power index $k_{ij}$ which only depends on the transformations of the fields $E^c_i$, $L_j$, $H_d$ under the modular generator $T$.
In particular the results are independent of the modular weight assignments of the fields. There exists basis in which the generators $S$ and $T$ are represented by unitary and symmetric matrices and the corresponding Clebsch-Gordan
coefficients are real, then the generalized CP symmetry reduces to the canonical CP and the all couplings are enforced to be real~\cite{Novichkov:2019sqv}. Then CP invariance requires the Yukawa couplings fulfill
\begin{equation}
\mathcal{Y}^e_{ij}(-\tau^{*})=\mathcal{Y}^{e*}_{ij}(\tau)\,,
\end{equation}
which follows from Eq.~\eqref{eq:me-mnu-CP}. Hence all the coefficients $C_{n}^{ij}$ are real due to CP symmetry, and the real part of $\tau$ is the unique source of CP violation. In the fundamental domain $\mathcal{D}$, the imaginary part $y=\text{Im}(\tau)$ is greater than or equal $\sqrt{3}/2$ so that the parameter $q$ is quite small,
\begin{eqnarray}
  |q|&=&e^{-2\pi y}\leq e^{-\sqrt{3}\pi}\simeq0.0043\ll1\,.
\end{eqnarray}
Hence the order of magnitude of the Yukawa couplings $\mathcal{Y}^e_{ij}(\tau)$ is dominated by the leading order term $q^{(N-k_{ij})/N}$.

\subsection{Weighton mechanism for further Yukawa suppressions}

We introduce a new chiral singlet superfield $\phi$ whose modular weight $k_{\phi}=1$. This $\phi$ is identified as a weighton, defined to be a complete singlet (under both finite modular symmetry and the SM gauge group) that carries one unit of modular weight \cite{King:2020qaj}. We also assume $\phi$ transforms under CP as a scalar. The weighton $\phi$ transforms under modular symmetry and CP as follows
\begin{eqnarray}
\phi&\stackrel{\gamma}{\longrightarrow}&
(c\tau+d)^{-1}\phi,~~~~\phi\stackrel{\mathcal{\mathcal{CP}}}{\longrightarrow}\phi^{\ast}\,,
\end{eqnarray}
After including weighton, the modular invariant Yukawa couplings become
\begin{equation}
\label{eq:Lag_mass-weighton}
\mathcal{W}=-\tilde{\phi}^{J_{ij}}E^{c}_i\mathcal{Y}^{e}_{ij}(\tau)L_jH_d-\frac{1}{2\Lambda}\tilde{\phi}^{K_{ij}}L_iL_j\mathcal{Y}^{\nu}_{ij}(\tau)H_uH_u\,,
\end{equation}
where $\tilde{\phi}=\phi/M_{fl}$ and $M_{fl}$ is cut-off flavour scale. It is straightforward to derive that the modular transformation of $\mathcal{Y}^{e}_{ij}(\tau)$ and $\mathcal{Y}^{\nu}_{ij}(\tau)$ is given by
\begin{eqnarray}
\nonumber&&\mathcal{Y}^e_{ij}\left(\gamma\tau\right)=(c\tau+d)^{k_{E^c_i}+k_{L_j}+k_d+J_{ij}}\rho^{*}_{d}(\gamma)\left[\rho^*_{E^c}(\gamma)\mathcal{Y}^e\left(\tau\right)\rho^{\dagger}_L(\gamma)\right]_{ij}\,,\,\\
\label{eq:me-mnu-gamma-weighton}&&\mathcal{Y}^\nu_{ij}\left(\gamma\tau\right)=(c\tau+d)^{k_{L_i}+k_{L_j}+2k_u+K_{ij}}\rho^{2*}_{u}(\gamma)\left[\rho^*_{L}(\gamma)\mathcal{Y}^\nu\left(\tau\right)\rho^{\dagger}_L(\gamma)\right]_{ij}\,.
\end{eqnarray}
By considering the $T$ transformation with $\gamma=T$, the same result as Eq.~\eqref{Yeij-expansion} is obtained, i.e.
\begin{eqnarray}
\label{Yeij-expansion2}\mathcal{Y}^e_{ij}(\tau)&=&q^{(N-k_{ij})/N}\left[C_{N-k_{ij}}^{ij}+C_{2N-k_{ij}}^{ij}q+C_{3N-k_{ij}}^{ij}q^2+\ldots\right]\,, \end{eqnarray}
with $\rho_{d}(T)\rho_{E^{c}_i}(T)\rho_{L_{j}}(T)=\zeta^{k_{ij}}$ in the $T$ diagonal basis. After the weighton gets a non-zero VEV, the effective Yukawa couplings are $\tilde{\phi}^{J_{ij}}\mathcal{Y}^{e}_{ij}(\tau)$ which are suppressed by powers of both $q$ and $\langle\tilde{\phi}\rangle=\langle\phi\rangle/M_{fl}$, its magnitude can be estimated as $\tilde{\phi}^{J_{ij}}q^{(N-k_{ij})/N}$ at leading order~\footnote{We will often use $\tilde{\phi}$ for its VEV $\langle\tilde{\phi}\rangle$ in the following.}.
Notice that, as discussed in~\cite{King:2020qaj}, higher order corrections suppressed by an additional two powers (or more even powers) of the weighton field are also present in general.

\section{\label{sec:hierarchical-patterns-Nleq5}Hierarchical patterns of fermion masses and mixing angles}

Using the general results in Eq.~\eqref{Yeij-expansion} and including the powers of weighton $\tilde{\phi}$, one can straightforwardly construct the hierarchical mass matrices once the representations $\bm{r}_{\psi}$ and $\bm{r}_{\psi^c}$ of the matter fields $\psi$ and $\psi^c$ are specified. The fermion mass matrix $M_{\psi}$ can be diagonalized by bi-unitary transformations as follows,
\begin{equation}
V^{T}_{\psi^c}M_{\psi}V_{\psi}=\text{diag}(m_{\psi_1}, m_{\psi_2}, m_{\psi_3})\,,\label{eq:VL_M_VR}
\end{equation}
which leads to
\begin{equation}
V^{\dagger}_{\psi}M^{\dagger}_{\psi}M_{\psi}V_{\psi}=\text{diag}\left(m^2_{\psi_1}, m^2_{\psi_2}, m^2_{\psi_3}\right)\,.
\end{equation}
The observed hierarchy of the quark and charged lepton masses implies that their Yukawa matrices have a hierarchical structure. We can permutate the rows and columns of $M_{\psi}$ to make the (33) entry being the largest and the off-diagonal entries being smaller than the diagonal entries, i.e.
\begin{equation}
M'_{\psi}=P_{\psi^c}M_{\psi}P^{T}_{\psi}\,,
\end{equation}
where $P_{\psi}$ and $P_{\psi^c}$ are some permutation matrices which can take the following six possible forms
\begin{eqnarray}
\nonumber&&P_{123}=\begin{pmatrix}
1    & 0     & 0   \,\\
0    & 1     & 0   \,\\
0    & 0     & 1
\end{pmatrix}\,,
~~
P_{231}=\begin{pmatrix}
0    & 1     & 0   \,\\
0    & 0     & 1   \,\\
1    & 0     & 0
\end{pmatrix} \,,
~~
P_{312}=\begin{pmatrix}
0    & 0     & 1   \,\\
1    & 0     & 0   \,\\
0    & 1     & 0
\end{pmatrix}\,,\\
\label{eq:permu-matr}&&
P_{132}=\begin{pmatrix}
1    & 0     & 0   \,\\
0    & 0     & 1   \,\\
0    & 1     & 0
\end{pmatrix}\,,
~~
P_{213}=\begin{pmatrix}
0    & 1     & 0   \,\\
1    & 0     & 0   \,\\
0    & 0     & 1
\end{pmatrix}\,,
~~
P_{321}=\begin{pmatrix}
0    & 0     & 1   \,\\
0    & 1     & 0   \,\\
1    & 0     & 0
\end{pmatrix}\,.
\end{eqnarray}
The mass matrix $M'_{\psi}$ can be perturbatively diagonalized by three successive rotations in the $(2, 3)$, $(1, 3)$ and $(1, 2)$ sectors,
\begin{equation}
V'^{T}_{\psi^c}M'_{\psi}V'_{\psi}=\text{diag}(m_{\psi_1}, m_{\psi_2}, m_{\psi_3})\,,
\end{equation}
with
\begin{eqnarray}
\nonumber && V'_{\psi}\simeq\begin{pmatrix}
1 & 0 & 0 \\
0 & 1 & s^{\psi}_{23} \\
0 & -(s^{\psi}_{23})^{*} & 1
\end{pmatrix} \begin{pmatrix}
1  & 0  & s^{\psi}_{13} \\
0  & 1  & 0\\
- (s^{\psi}_{13})^{*} & 0 & 1
\end{pmatrix} \begin{pmatrix}
1  &  s^{\psi}_{12}  & 0 \\
- (s^{\psi}_{12})^{*} & 1  & 0\\
0 & 0  & 1
\end{pmatrix}\,,\\
&& V'_{\psi^c}\simeq\begin{pmatrix}
1 & 0 & 0 \\
0 & 1 & s^{\psi^c}_{23} \\
0 & -(s^{\psi^c}_{23})^{*} & 1
\end{pmatrix} \begin{pmatrix}
1  & 0  & s^{\psi^c}_{13} \\
0  & 1  & 0\\
- (s^{\psi^c}_{13})^{*} & 0 & 1
\end{pmatrix} \begin{pmatrix}
1  &  s^{\psi^c}_{12}  & 0 \\
-(s^{\psi^c}_{12})^{*} & 1  & 0\\
0 & 0  & 1
\end{pmatrix}\,.
\end{eqnarray}
Here the diagonalization parameters $s^{\psi}_{ij}$ and $s^{\psi^c}_{ij}$ can be expressed in terms of the elements of the mass matrix $M_{\psi}$, and they can be parametrized by rotation angles $\theta^{\psi}_{ij}$ and $\theta^{\psi^c}_{ij}$ as $|s^{\psi}_{ij}|=\sin\theta^{\psi}_{ij}$ and $|s^{\psi^c}_{ij}|=\sin\theta^{\psi^{c}}_{ij}$ respectively. Thus the unitary transformations $V_{\psi}$ and $V_{\psi^c}$ are given by
\begin{equation}
V_{\psi}=P^{T}_{\psi}V'_{\psi},~~~V_{\psi^c}=P^{T}_{\psi^c}V'_{\psi^c}\,.
\end{equation}
Since the unitary rotation of the electroweak left-handed fields enters into the charged current interactions in SM, only $V_{\psi}$ is relevant.

The general results of our analysis are presented in the lengthy but comprehensive Appendix~\ref{app:patterns-masses}.
For different representation and modular weight assignments of $\psi$ and $\psi^c$ under the finite modular group $\Gamma'_N$, the hierarchical patterns of the fermion masses and rotation angles $\theta^{\psi}_{ij}$ as well as $P_{\psi}$ are summarized in table~\ref{tab:A4p-patterns}, table~\ref{tab:S4p-patterns} and table~\ref{tab:A5p-patterns} for $N=3$, $N=4$ and $N=5$ respectively~\footnote{Notice that the accidental cancellation in certain models could spoil the general results in tables~\ref{tab:A4p-patterns}, ~\ref{tab:S4p-patterns} and ~\ref{tab:A5p-patterns}.}. For these results, the three generations of matter fields $\psi\equiv(\psi_1, \psi_2, \psi_3)^T$ and $\psi^{c}\equiv(\psi^c_1, \psi^c_2, \psi^c_3)^T$ are assigned to three-dimensional (possible reducible) representations $\bm{r}_{\psi}$ and $\bm{r}_{\psi^c}$ of $\Gamma'_N$ respectively. We require at least one of $\bm{r}_{\psi}$ and $\bm{r}_{\psi^c}$ to be an irreducible triplet of $\Gamma'_N$, which is part of the motivation for using non-Abelian finite discrete symmetry for addressing the flavour problem, and also for the practical reason that otherwise there are too many cases to consider after including the weighton. As a consequence, the superpotential for the mass of $\psi$ involves at most three power indices of $\tilde{\phi}$. For instance, if the $\psi$ is a triplet and $\psi^c_1$, $\psi^c_2$ and $\psi^c_3$ are three singlets of $\Gamma'_N$, the superpotential can be generally written as,
\begin{eqnarray}
\label{eq:spp-IJK}
\mathcal{W}=\alpha\;\tilde{\phi}^I\left(\psi^c_1 \psi Y^{(k_1)}_{\bm{r}_1} H_{u/d}\right)_{\bm{1}}+\beta\;\tilde{\phi}^J\left(\psi^c_2 \psi Y^{(k_2)}_{\bm{r}_2} H_{u/d}\right)_{\bm{1}}+\gamma\;\tilde{\phi}^{K}\left(\psi^c_3 \psi Y^{(k_3)}_{\bm{r}_3} H_{u/d}\right)_{\bm{1}}\,,
\end{eqnarray}
where the power indices $I$, $J$ and $K$ are are determined by the modular weights of the matter fields $\psi$, $\psi^c$ and Higgs field $H_{u/d}$. Notice that one should consider the contributions of all possible modular forms compatible with modular invariance and all independent singlet contractions should be summed over. If the first two generation fields $\psi^c_1$ and $\psi^c_2$ are assigned to a doublet of $\Gamma'_N$, we have $I=J$. If $\psi^c_1$, $\psi^c_2$ and $\psi^c_3$ are embedded into an irreducible triplet, then we have $I=J=K$. The superpotential has a similar form as Eq.~\eqref{eq:spp-IJK} in the case that $\psi^c$ instead of $\psi$ is an irreducible triplet of $\Gamma'_N$.

As an example of the application of the general results in Appendix~\ref{app:patterns-masses}, let us consider a modular invariant model at level $N=3$ with matter fields in the representations $\bm{r}_{\psi}=\bm{3}$ and $\bm{r}_{\psi^c}=\bm{1}^{\prime\prime}\oplus\bm{1}^{\prime}\oplus\bm{1}$ which means $\bm{r}_{\psi^c_1}=\bm{1}^{\prime\prime}$, $\bm{r}_{\psi^c_2}=\bm{1}^{\prime}$, $\bm{r}_{\psi^c_3}=\bm{1}$. Then the leading order superpotential of the Yukawa couplings takes the following form,
\begin{eqnarray}
\mathcal{W}=\alpha\;\tilde{\phi}^I\psi^c_1 \left(\psi Y^{(k_1)}_{\bm{3}}\right)_{\bm{1'}} H_{u/d}+\beta\;\tilde{\phi}^J\psi^c_2 \left(\psi Y^{(k_2)}_{\bm{3}}\right)_{\bm{1''}} H_{u/d}+\gamma\;\tilde{\phi}^{K}\psi^c_3\left(\psi Y^{(k_3)}_{\bm{3}}\right)_{\bm{1}} H_{u/d}\,.
\end{eqnarray}
We can straightforwardly read out the fermion mass matrix,
\begin{equation}
M_{\psi}=\frac{1}{\sqrt{3}}\begin{pmatrix}
\alpha\tilde{\phi}^I Y^{(k_1)}_{\bm{3}, 2}  & \alpha\tilde{\phi}^I Y^{(k_1)}_{\bm{3}, 1}  &  \alpha\tilde{\phi}^I Y^{(k_1)}_{\bm{3}, 3} \\
\beta\tilde{\phi}^J Y^{(k_2)}_{\bm{3}, 3}  & \beta\tilde{\phi}^J Y^{(k_2)}_{\bm{3}, 2}  &  \beta\tilde{\phi}^J Y^{(k_2)}_{\bm{3}, 1} \\
\gamma\tilde{\phi}^K Y^{(k_3)}_{\bm{3}, 1}  &  \gamma\tilde{\phi}^K Y^{(k_3)}_{\bm{3}, 3}   &  \gamma\tilde{\phi}^K Y^{(k_3)}_{\bm{3}, 2}
\end{pmatrix}v_{u/d}\,.
\end{equation}
By considering the $q$-expansions in Appendix~\ref{subsec:modular-forms-level-3}, we see that the leading terms of three components of a level $N=3$ triplet modular form $Y^{(k)}_{\bm{3}}$ are generally of order one, $q^{1/3}$ and $q^{2/3}$ respectively, i.e.
\begin{eqnarray}
Y^{(k)}_{\bm{3}, 1}\sim\mathcal{O}(1)\,,~~~Y^{(k)}_{\bm{3}, 2}\sim\mathcal{O}(q^{1/3})\,,~~~Y^{(k)}_{\bm{3}, 3}\sim\mathcal{O}(q^{2/3})\,.
\end{eqnarray}
Hence the mass matrix $M_{\psi}$ can be approximated by
\begin{equation}
M_{\psi}\sim\begin{pmatrix}
\alpha\tilde{\phi}^I q^{1/3}  & \alpha\tilde{\phi}^I   &  \alpha\tilde{\phi}^I q^{2/3} \\
\beta\tilde{\phi}^J q^{2/3} & \beta\tilde{\phi}^J q^{1/3}  &  \beta\tilde{\phi}^J  \\
\gamma\tilde{\phi}^K   &  \gamma\tilde{\phi}^K q^{2/3}   &  \gamma\tilde{\phi}^K q^{1/3}
\end{pmatrix}v_{u/d}\,.
\label{Mapprox}
\end{equation}
The mass matrix approximation in Eq.~\eqref{Mapprox} can also readily be obtained from the general result in Eq.~\eqref{Yeij-expansion}. For example, from the representation matrices of $\rho_{\bm{r}_{\psi}}(T)$ and $\rho_{\bm{r}_{\psi^c}}(T)$ in table~\ref{tab:representation-matrices-Tp} we can read out $k_{11}=2$ for the (11) element. Then using Eq.~\eqref{Yeij-expansion2} we can straightforwardly obtain its order of magnitude as $\alpha\tilde{\phi}^I q^{1/3}$.

For the case of weighton power indices $K>I>J$, the permutation matrices can be taken as $P_{\psi^c}=P_{312}$ and $P_{\psi}=P_{123}$. The resulting mass matrix can be perturbatively diagonalized with the rotation angles
\begin{equation}
\label{eq:angles-example}\theta^{\psi}_{12}\sim|q|^{1/3}\,,~~~\theta^{\psi}_{23}\sim|q|^{1/3}\,,~~~\theta^{\psi}_{13}\sim(|q|^{1/3}\tilde{\phi}^{2(K-J)}+|q|^{2/3})\,.
\end{equation}
One can also obtain the approximate expressions of the fermion masses as follows,
\begin{equation}
\label{eq:masses-example}m_{\psi_1}\sim\tilde{\phi}^K\,,~~~m_{\psi_2}\sim\tilde{\phi}^I\,,~~~m_{\psi_3}\sim\tilde{\phi}^J\,.
\end{equation}
In this example the $q$-expansion plays a role in the mixing angles, while the weighton expansion controls the fermion mass hierarchy. Note that there could be more than one linearly independent triplet modular forms at higher weights $k_1$, $k_2$, $k_3$, as can be seen from table~\ref{tab:modular-forms-level-3}. Then the contributions of the all-relevant triplet modular forms should be taken into account, the estimates in Eqs.~(\ref{eq:angles-example}, \ref{eq:masses-example}) still hold true except that certain accidental cancellation occurs.

As regards the neutrino sector, the light neutrino masses are assumed to be generated by the Weinberg operator, and as usual the three generations of lepton doublets are assigned to irreducible triplets of $\Gamma'_N$ in order to account for the large lepton mixing. Thus the modular invariant neutrino mass matrix is of the following form
\begin{equation}
M_{\nu}\sim\begin{pmatrix}
\times  & 0 & 0 \\
0  & 0 & \times \\
0  & \times & 0
\end{pmatrix}\,~~\text{or}~~
\begin{pmatrix}
0 & 0 & 0 \\
0  & \times & 0 \\
0  & 0 & \times
\end{pmatrix}
\end{equation}
at leading order, where ``$\times$'' denotes non-vanishing entry. The light neutrino mass matrix can be diagonalized as
\begin{equation}
V^{T}_{\nu}M_{\nu}V_{\nu}=\text{diag}(m_{\nu_1}, m_{\nu_2}, m_{\nu_3})\,,
\end{equation}
with
\begin{eqnarray}
\nonumber &&  V_{\nu}=P^{T}_{\nu}V'_{\nu}\,,~~P_{\nu}=O_{23}=\frac{1}{\sqrt{2}}\left( \begin{array}{ccc}
\sqrt{2}    & 0     & 0   \,\\
0          & 1     & -1   \,\\
0          & 1    & 1
\end{array} \right)~~\text{or}~~P_{\nu}=P_{123}\,,\\
&&V'_{\nu}\simeq\begin{pmatrix}
1 & 0 & 0 \\
0 & 1 & s^{\nu}_{23} \\
0 & -(s^{\nu}_{23})^{*} & 1
\end{pmatrix} \begin{pmatrix}
1  & 0  & s^{\nu}_{13} \\
0  & 1  & 0\\
- (s^{\nu}_{13})^{*} & 0 & 1
\end{pmatrix} \begin{pmatrix}
1  &  s^{\nu}_{12}  & 0 \\
- (s^{\nu}_{12})^{*} & 1  & 0\\
0 & 0  & 1
\end{pmatrix}\,.
\end{eqnarray}
The order of magnitudes of the light neutrino masses and the rotation angles $\theta^{\nu}_{ij}$ are listed in table~\ref{tab:neutrino-patterns}.

\section{\label{sec:example-models}Example models}

In the following, we present two example models based the finite modular group $\Gamma'_3\cong T'$, which can explain the measured masses and mixing of both quarks and leptons simultaneously, while also giving considerable insight into the respective hierarchies. The neutrino masses are described by the effective Weinberg operator in the model A and the best fit value of the modulus $\tau$ is close to the pure imaginary axis. In the second model B, the light neutrino masses are generated by the type-I seesaw mechanism, and the best fit value of $\tau$ is in the vicinity of the left vertical boundary of the fundamental domain.

\subsection{Model A}

\begin{figure}[t!]
\begin{center}
\includegraphics[width=0.48\textwidth]{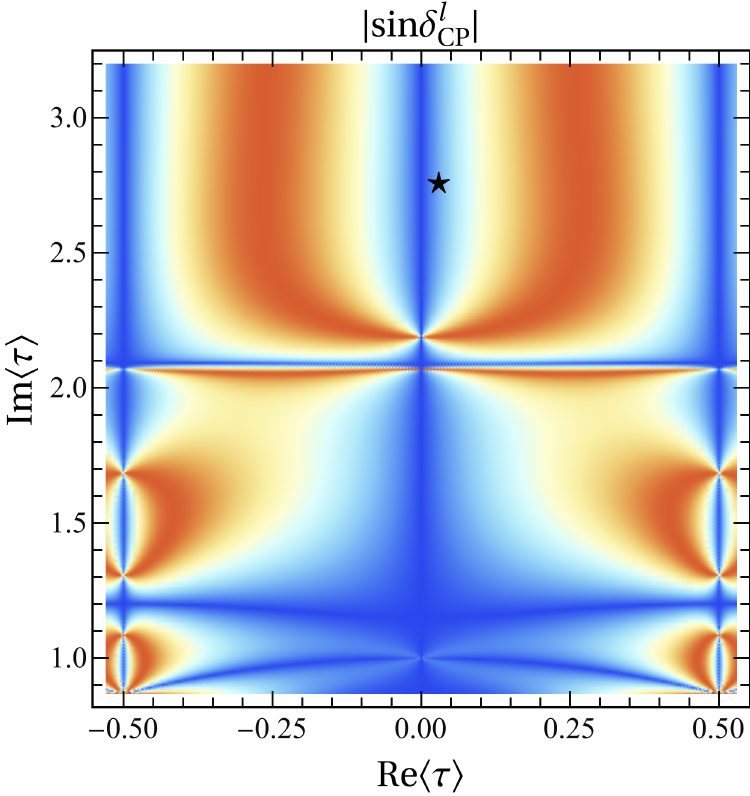}  \includegraphics[width=0.48\textwidth]{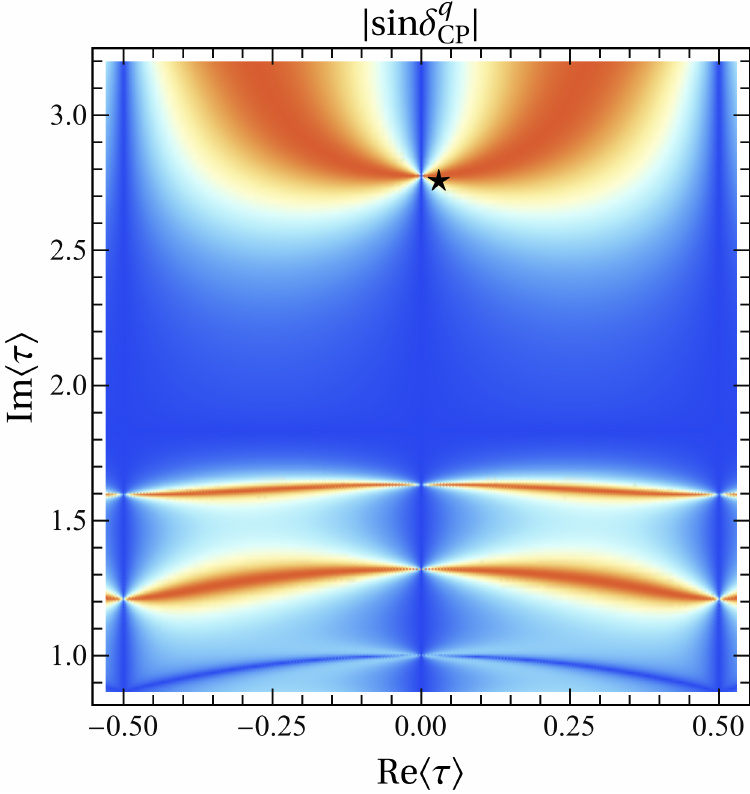} \\
\includegraphics[width=0.48\textwidth]{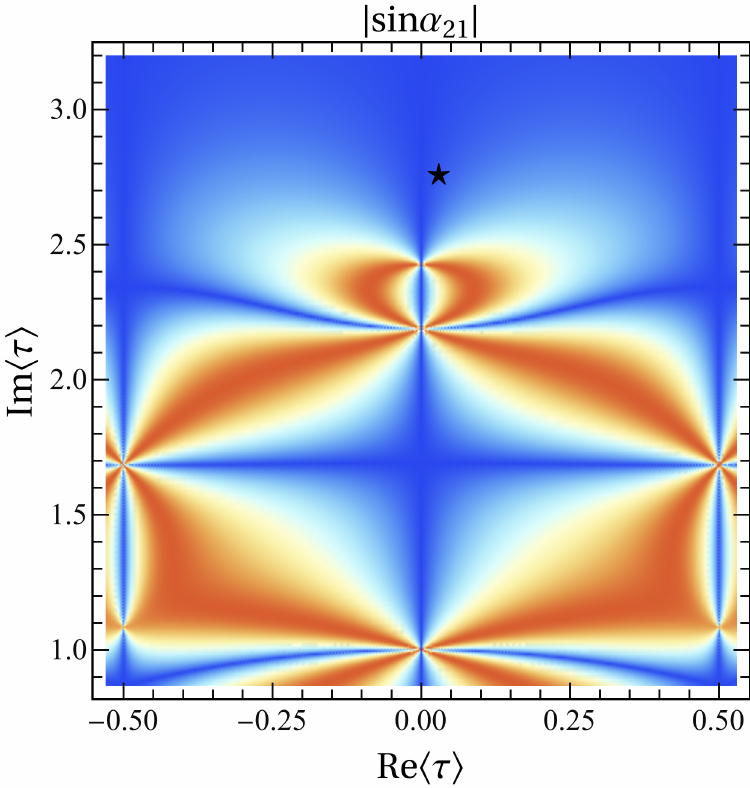} \includegraphics[width=0.48\textwidth]{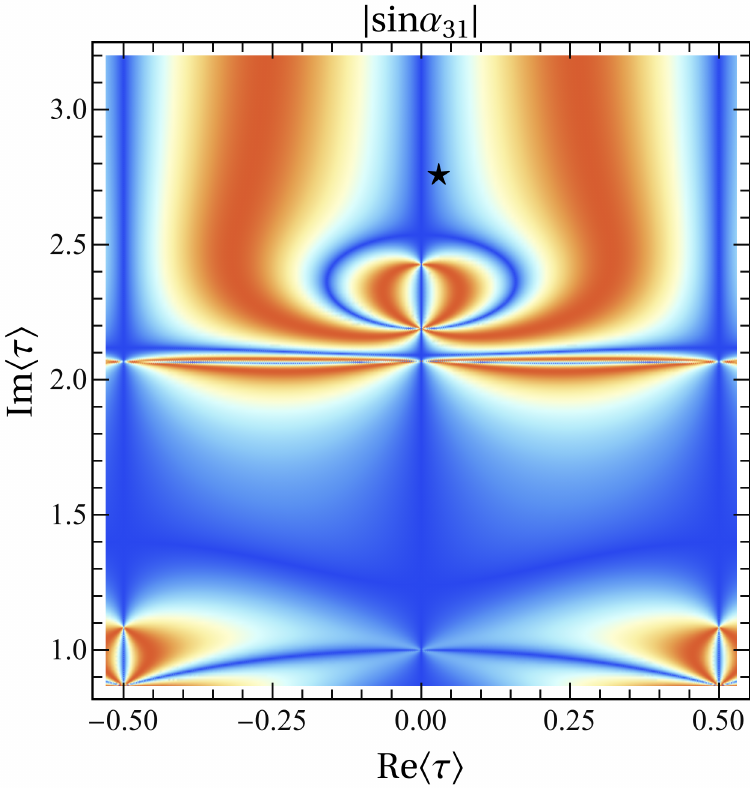}\\
\includegraphics[width=0.8\textwidth]{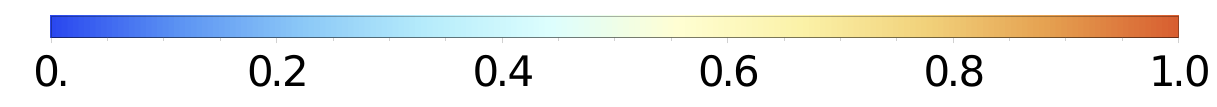}
\end{center}
\caption{\label{fig:contour-model-A}The contour of $|\sin\delta^{l}_{CP}|$, $|\sin\delta^{q}_{CP}|$, $|\sin\alpha_{21}|$ and $|\sin\alpha_{31}|$ in the plane of $\tau$ for the model A, where black star refers to the best fitting point of $\tau$. The couplings are set to the best fit values in Eq.~\eqref{eq:bf-pars-model-A}}
\end{figure}

The generalized CP symmetry is imposed and it constrains all coupling constants to be real in our basis. We assign the three generations of left-handed lepton doublet $L$ and quark doublet $Q_L$ to two irreducible triplets $\bm{3}$ of $T'$,  the right-handed charged leptons $e^c$, $\mu^c$, $\tau^c$ and right-handed up type quarks $u^c$, $c^c$, $t^c$ are singlets of $T'$, while the right-handed down type quarks $d^c$ and $s^c$ are assigned to a doublet $\bm{\widehat{2}}^{\prime}$ of $T'$, and $b^c$ is invariant under $T'$. The transformations of the matter fields under the action of modular symmetry are given by,
\begin{eqnarray}\nonumber
&&L\sim \left(\bm{3}, 2\right)\,,~~e^{c}\sim \left(\bm{1}, -1\right)\,,~~\mu^{c}\sim \left(\bm{1}, 1\right)\,,~~\tau^{c}\sim\left(\bm{1''}, 3\right)\,,~~H_{u,d}\sim\left(\bm{1}, 0\right)\,, \,\\
\nonumber&&Q_{L}\sim\left(\bm{3}, k_{Q_{L}}\right)\,,~~u^{c}\sim\left(\bm{1}, -k_{Q_{L}}\right)\,,~~c^{c}\sim\left(\bm{1''}, 5-k_{Q_{L}}\right)\,,~~t^{c}\sim\left(\bm{1'}, 4-k_{Q_{L}}\right)\,,\,\\
&&D_{D}^{c}\equiv\{d^{c},s^{c}\}\sim\left(\bm{\widehat{2}'}, 2-k_{Q_{L}}\right)\,,~~b^{c}\sim \left(\bm{1}, 6-k_{Q_{L}}\right)\,,~~\phi\sim\left(\bm{1}, 1\right)\,.
\end{eqnarray}
Here the modular transformation of a field is denoted as $\psi\sim\left(\bm{r}, k_{\psi}\right)$, where $\bm{r}$ stands for the transformation under the finite modular group $\Gamma'_3\cong T'$ and $k_{\psi}$ refers to its modular weight. Then we can read off the modular invariant mass terms for leptons and quarks as follow,
\begin{eqnarray}\nonumber
\nonumber\mathcal{W}_{e}&=&y^{e}_{1}\tilde{\phi}e^{c}\left(LY^{(2)}_{\bm{3}}\right)_{\bm{1}}H_{d}+y^{e}_{2}\tilde{\phi}\mu^{c}\left(LY^{(4)}_{\bm{3}}\right)_{\bm{1}}H_{d}+ y^{e}_{3}\tau^{c}\left(LY^{(6)}_{\bm{3}A}\right)_{\bm{1'}}H_{d} + y^{e}_{4}\tau^{c}\left(LY^{(6)}_{\bm{3}B}\right)_{\bm{1'}}H_{d}\,,\,\\
\nonumber\mathcal{W}_{\nu}&=&\frac{y^{\nu}_{1}}{\Lambda}\left((LL)_{\bm{1}}Y^{(4)}_{\bm{1}}\right)_{\bm{1}}H_{u}H_{u}+\frac{y^{\nu}_{2}}{\Lambda}\left((LL)_{\bm{1''}}Y^{(4)}_{\bm{1'}}\right)_{\bm{1}}H_{u}H_{u}
+\frac{y^{\nu}_{3}}{\Lambda}\left((LL)_{\bm{3}_{S}}Y^{(4)}_{\bm{3}}\right)_{\bm{1}}H_{u}H_{u}\,,\\
\nonumber\mathcal{W}_{u}&=&y^{u}_{1}\tilde{\phi}^{2}
u^{c}(Q_{L}Y^{(2)}_{\bm{3}})_{\bm{1}}H_{u}
+y^{u}_{2}\tilde{\phi}c^{c}
(Q_{L}Y^{(6)}_{\bm{3}A})_{\bm{1'}}H_{u}
+y^{u}_{3}\tilde{\phi}c^{c}(Q_{L}Y^{(6)}_{\bm{3}B})_{\bm{1'}}
H_{u}+y^{u}_{4}t^{c}(Q_{L}Y^{(4)}_{\bm{3}})_{\bm{1''}}
H_{u}\nonumber\,,\,\\
\nonumber\mathcal{W}_{d}&=&y^{d}_{1}\tilde{\phi}
\left((D_{D}^{c}Q_{L})_{\bm{\widehat{2}}}Y^{(3)}_{\bm{\widehat{2}''}}
\right)_{\bm{1}}
H_{d}+y^{d}_{2}\tilde{\phi}
\left((D_{D}^{c}Q_{L})_{\bm{\widehat{2}''}}
Y^{(3)}_{\bm{\widehat{2}}}\right)_{\bm{1}}H_{d}+y^{d}_{3}
b^{c}
(Q_{L}Y^{(6)}_{\bm{3}A})_{\bm{1}}H_{d} \nonumber\,\\
&&+ y^{d}_{4}b^{c}(Q_{L}Y^{(6)}_{\bm{3}B})_{\bm{1}}H_{d}\,.
\end{eqnarray}
This model is very predictive, and it uses 18 real free parameters to describe the 22 masses and mixing parameters of quarks and leptons. Here we have counted $\tilde{\phi}$ as a free parameter. Using the Clebsch–Gordan coefficients in Appendix~\ref{app:Tp-group} and expanding the above superpotential, we can obtain the quark and lepton mass matrices as
\begin{eqnarray}
\nonumber M_{e}&=&\frac{v_{d}}{\sqrt{3}}\left(
\begin{array}{ccc}
y^{e}_{1} \tilde{\phi}Y_{\bm{3},1}^{(2)} & y^{e}_{1} \tilde{\phi}Y_{\bm{3},3}^{(2)}  & y^{e}_{1} \tilde{\phi}Y_{\bm{3},2}^{(2)} \,\\
 y^{e}_{2} \tilde{\phi}Y_{\bm{3},1}^{(4)}  & y^{e}_{2} \tilde{\phi}Y_{\bm{3},3}^{(4)}  & y^{e}_{2} \tilde{\phi}Y_{\bm{3},2}^{(4)} \,\\
 y^{e}_{3} Y_{\bm{3}A,2}^{(6)} +y^{e}_{4} Y_{\bm{3}B,2}^{(6)} & y^{e}_{3} Y_{\bm{3}A,1}^{(6)}+y^{e}_{4} Y_{\bm{3}B,1}^{(6)} & y^{e}_{3} Y_{\bm{3}A,3}^{(6)} +y^{e}_{4} Y_{\bm{3}B,3}^{(6)}\,\\
\end{array}
\right)\,,\,\\
\nonumber M_{\nu} &=&\frac{v_{u}^{2}}{3\sqrt{2}\Lambda} \begin{pmatrix}
  \sqrt{6}y^{\nu}_{1}Y^{(4)}_{\bm{1}}+2y^{\nu}_{3}Y^{(4)}_{\bm{3},1} ~&~ -y^{\nu}_{3}Y^{(4)}_{\bm{3},3} ~&~ \sqrt{6}y^{\nu}_{2}Y^{(4)}_{\bm{1}^{\prime}}-y^{\nu}_{3}Y^{(4)}_{\bm{3},2} \,\\
 -y^{\nu}_{3}Y^{(4)}_{\bm{3},3} ~&~ \sqrt{6}y^{\nu}_{2}Y^{(4)}_{\bm{1}^{\prime}}+2y^{\nu}_{3}Y^{(4)}_{\bm{3},2}  ~&~ \sqrt{6}y^{\nu}_{1}Y^{(4)}_{\bm{1}}-y^{\nu}_{3}Y^{(4)}_{\bm{3},1}  \,\\
 \sqrt{6}y^{\nu}_{2}Y^{(4)}_{\bm{1}^{\prime}}-y^{\nu}_{3}Y^{(4)}_{\bm{3},2} ~&~ \sqrt{6}y^{\nu}_{1}Y^{(4)}_{\bm{1}}-y^{\nu}_{3}Y^{(4)}_{\bm{3},1} ~&~2y^{\nu}_{3}Y^{(4)}_{\bm{3},3} \,\\
\end{pmatrix}\,,\\
 \nonumber M_{u}&=&\frac{v_{u}}{\sqrt{3}}
 \left( \begin{array}{ccc}
 y^{u}_{1}\tilde{\phi}^{2}Y^{(2)}_{\bm{3},1} ~& y^{u}_{1}\tilde{\phi}^{2}Y^{(2)}_{\bm{3},3} &~ y^{u}_{1}\tilde{\phi}^{2}Y^{(2)}_{\bm{3},2}  \,\\
y^{u}_{2}\tilde{\phi}Y^{(6)}_{\bm{3}A,2}+y^{u}_{3}\tilde{\phi}Y^{(6)}_{\bm{3}B,2} ~& y^{u}_{2}\tilde{\phi}Y^{(6)}_{\bm{3}A,1}+y^{u}_{3}\tilde{\phi}Y^{(6)}_{\bm{3}B,1}  &~ y^{u}_{2}\tilde{\phi}Y^{(6)}_{\bm{3}A,3}+y^{u}_{3}\tilde{\phi}Y^{(6)}_{\bm{3}B,3}   \,\\
y^{u}_{4}Y^{(4)}_{\bm{3},3} ~& y^{u}_{4}Y^{(4)}_{\bm{3},2} &~ y^{u}_{4}Y^{(4)}_{\bm{3},1}  \,\\ \end{array} \right)\,,\,\\
   M_{d}&=&\frac{v_{d}}{\sqrt{6}}\left(
\begin{array}{ccc}
 \sqrt{2}y^{d}_{2}\tilde{\phi} Y_{\bm{\widehat{2}},1}^{(3)}
  &
  -y^{d}_{2}\tilde{\phi} Y_{\bm{\widehat{2}},2}^{(3)}
  -\sqrt{2}y^{d}_{1}\tilde{\phi}Y_{\bm{\widehat{2}}^{\prime\prime},1}^{(3)}
  &
  y^{d}_{1}\tilde{\phi}Y_{\bm{\widehat{2}}^{\prime\prime},2}^{(3)}
  \,\\
 \sqrt{2}y^{d}_{1}\tilde{\phi}Y_{\bm{\widehat{2}}^{\prime\prime},2}^{(3)}
 &
  -y^{d}_{2}\tilde{\phi} Y_{\bm{\widehat{2}},1}^{(3)}
  &
   y^{d}_{1}\tilde{\phi}Y_{\bm{\widehat{2}}^{\prime\prime},1}^{(3)}
 -\sqrt{2}y^{d}_{2}\tilde{\phi} Y_{\bm{\widehat{2}},2}^{(3)}
 \,\\
 \sqrt{2}y^{d}_{3} Y_{\bm{3}A,1}^{(6)}
 +\sqrt{2}y^{d}_{4} Y_{\bm{3}B,1}^{(6)}
 &
  \sqrt{2}y^{d}_{3} Y_{\bm{3}A,3}^{(6)}
 +\sqrt{2}y^{d}_{4} Y_{\bm{3}B,3}^{(6)}
 &
  \sqrt{2}y^{d}_{3} Y_{\bm{3}A,2}^{(6)}
  +\sqrt{2}y^{d}_{4} Y_{\bm{3}B,2}^{(6)} \nonumber
  \,\\
\end{array}
\right) \,.\\
\end{eqnarray}
In order to quantitatively assess how well the model can accommodate the experimental observations, we perform a $\chi^2$ analysis to determine the best-fit values of the model parameters and the corresponding predictions for fermion masses and mixing parameters. We look for the minimum of the $\chi^2$ built with the following experimental data and errors,
\begin{equation}
\label{eq:mixing_data}
\begin{array}{c}
\sin ^{2} \theta^{l}_{12}=0.308_{-0.011}^{+0.012}\,, \quad \sin ^{2} \theta^{l}_{13}=0.02215_{-0.00058}^{+0.00056}\,, \quad \sin ^{2} \theta^{l}_{23}=0.470_{-0.013}^{+0.017}\,, \\
\delta_{C P}^{l} /^{\circ}=212_{-41}^{+26}\,, \quad \frac{\Delta m_{21}^{2}}{10^{-5} \mathrm{eV}^{2}}=7.49_{-0.19}^{+0.19}\,, \quad \frac{\Delta m_{31}^{2}}{10^{-3} \mathrm{eV}^{2}}=2.513_{-0.019}^{+0.021}\,,\\
m_{e}/m_{\mu}=0.0048 \pm 0.0002 , \quad m_{\mu}/m_{\tau}=0.059 \pm 0.002\,,\quad m_{\tau}/\text{GeV}=1.219\pm 0.052 \,,\\
m_{u} / m_{c}=0.0027 \pm 0.0006\,, \quad m_{c} / m_{t}=0.0025\pm 0.0002\,,\quad m_{t}/\text{GeV}=83.155\pm 3.465\,,\\
m_{d} / m_{s}=0.051 \pm 0.007\,, \quad m_{s} / m_{b}= 0.019\pm 0.002\,,\quad m_{b}/\text{GeV}=0.884\pm 0.035\,, \\
\theta_{12}^{q}=0.229 \pm 0.001\,, \quad \theta_{13}^{q}=0.0037 \pm 0.0004\,,\quad \theta_{23}^{q}=0.0397 \pm 0.0011\,,\quad \delta_{C P}^{q}/^{\circ}=56.34 \pm 7.89\,.
\end{array}
\end{equation}
Here, the values of the lepton mixing angles $\theta^{l}_{12}, \theta^{l}_{13}, \theta^{l}_{23}$, the leptonic Dirac CP phase $\delta^{l}_{CP}$, and the neutrino mass-squared differences are taken from NuFIT v6.0 with Super-Kamiokande atmospheric data for normal ordering neutrino masses~\cite{Esteban:2024eli}. The central values and $1\sigma$ uncertainties of the charged lepton masses, quark masses, and quark mixing parameters at the GUT scale are adopted from Ref.~\cite{Ross:2007az}, assuming $\tan\beta = 10$ and a supersymmetry breaking scale of $M_{\text{SUSY}} = 500$~GeV. We get a good agreement between the model and the data for the following parameter choice
\begin{eqnarray}
\nonumber&&\langle\tau\rangle=0.03055 + 2.7572i,~~
\langle\tilde{\phi}\rangle=0.002,\\ \nonumber
&&y^{e}_{1}/y^{e}_{3}=0.1350,
~~y^{e}_{2}/y^{e}_{3}=2.7065\,,
~~y^{e}_{4}/y^{e}_{3}=1.8828,
~~y^{e}_{3}v_{d}=57.2137\text{GeV}\,,\\ \nonumber
&&y^{\nu}_{2}/y^{\nu}_{1}=12.8964,
~~y^{\nu}_{3}/y^{\nu}_{1}=3.5915,
~~(y^{\nu}_{1}v_{u})^{2}/\Lambda=37.2464\; \text{meV}\,,\,\\
\nonumber &&y^{u}_{1}/y^{u}_{4}=1.1381,
~~y^{u}_{2}/y^{u}_{4}=11.4990,
~~y^{u}_{3}/y^{u}_{4}=24.4417,
~~y^{u}_{4}v_{u}=249.4223~\text{GeV}\,,\,\\
\label{eq:bf-pars-model-A}&&y^{d}_{1}/y^{d}_{3}=0.5442,
~~y^{d}_{2}/y^{d}_{3}=1.5303,
~~y^{d}_{4}/y^{d}_{3}=1.7297,
~~y^{d}_{3}v_{d}=61.4147~\text{GeV}\,,
\end{eqnarray}
with the minimum $\chi^2_{\text{min}}=16.8$. The masses and mixing parameters of quarks and leptons at the above best fitting point are determined to be,
\begin{eqnarray}
\nonumber&& \sin^{2}\theta^{l}_{12}=0.283\,,\quad \sin^{2}\theta^{l}_{13}=0.02136\,,\quad \sin^{2}\theta^{l}_{23}=0.470\,,\quad \delta^{l}_{CP}=189.9^{\circ}\,,\,\\
\nonumber&& \alpha_{21}=0.991\pi\,,\quad \alpha_{31}=0.0461\pi\,,\quad m_e/m_{\mu}=0.004681\,,\quad m_{\mu}/m_{\tau}=0.05795\,,\,\\
\nonumber&&  m_1=3.14~\text{meV}\,,\quad m_2=9.21~\text{meV}\,,\quad m_3=49.63~\text{meV}\,,\quad m_{\beta\beta}=0.71~\text{meV}\,,\,\\
\nonumber&& \theta^q_{12}=0.2291\,,\quad \theta^q_{13}=0.003390\,,\quad \theta^q_{23}=0.03975\,,\quad \delta^q_{CP}=54.61^{\circ} \,,\,\\
&& m_u / m_c=0.002488\,,\quad m_c / m_t=0.002877\,,\quad m_d/m_s=0.04965\,,\quad m_s / m_b=0.01526\,.
\end{eqnarray}
We see that all flavor observables are predicted to lie in the experimentally preferred $\bm{3}\sigma$ ranges. Note that $m_{\beta\beta}$ is the effective neutrino mass in neutrinoless double beta decay, it is much below the present upper limit $m_{\beta\beta} < (28- 122)~\text{meV}$ from KamLAND-Zen collaboration ~\cite{KamLAND-Zen:2024eml} and the sensitivities of the future neutrinoless double beta decay experiments~\cite{Agostini:2022zub} in this model. It is remarkable that the VEV of $\tau$ is very close to the pure imaginary axis. If one sets $\langle\tau\rangle=2.7572i$ and the values of the other parameters are those in Eq.~\eqref{eq:bf-pars-model-A}, the lepton and quark mass matrices would be real so that all the CP phases are conserved, while the predictions for the masses and mixing angles of quarks and leptons almost don't change, i.e.
\begin{eqnarray}
\nonumber&& \sin^{2}\theta^{l}_{12}=0.283\,,\quad \sin^{2}\theta^{l}_{13}=0.02136\,,\quad \sin^{2}\theta^{l}_{23}=0.470\,,\quad \delta^{l}_{CP}=180^{\circ}\,,\,\\
\nonumber&& \alpha_{21}=\pi\,,\quad \alpha_{31}=0\,,\quad m_e/m_{\mu}=0.004681\,,\quad m_{\mu}/m_{\tau}=0.05795\,,\quad \frac{\Delta m_{21}^{2}}{\Delta m_{31}^{2}} = 0.03058\,,\\
\nonumber&&  m_1=3.12788~\text{meV}\,,\quad m_2=9.20237~\text{meV}\,,\quad m_3=49.5896~\text{meV}\,,\quad m_{\beta\beta}=0.70928~\text{meV}\,,\,\\
\nonumber&& \theta^q_{12}=0.2286\,,\quad \theta^q_{13}=0.001729\,,\quad \theta^q_{23}=0.03986\,,\quad \delta^q_{CP}=0^{\circ} \,,\,\\
&& m_u / m_c=0.002488\,,\quad m_c / m_t=0.002877\,,\quad m_d/m_s=0.04797\,,\quad m_s / m_b=0.01526\,.~~~~~
\end{eqnarray}
Therefore the role of small $\text{Re}(\tau)$ is to generate the CP violation phases. We display the contour plot of  $|\sin\delta^{l}_{CP}|$, $|\sin\delta^{q}_{CP}|$, $|\sin\alpha_{21}|$ and $|\sin\alpha_{31}|$ in the complex plane of $\tau$ in figure~\ref{fig:contour-model-A}. One sees that the CP violation phases are very sensitive to $\tau$, a small deviation from the CP conserved lines can lead to large CP violation, while the successful predictions for the fermion masses and mixing angles are preserved.

We proceed to give an analytical estimate in the following. From Eq.~\eqref{eq:Y1-Y2-q-exp}, we see that the weight one modular forms $Y_1(\tau)$
and $Y_2(\tau)$ can be approximated as
\begin{equation}
Y_1(\tau)=3\sqrt{2}q^{1/3}+\mathcal{O}(q^{4/3})\equiv\eta\,,~~Y_{2}(\tau)=-1+\mathcal{O}(q)\,,
\end{equation}
with $\eta\approx0.0132$ at the best fit value of $\tau$. Thus the leading order approximation for the charged lepton and neutrino mass matrices is given by
\begin{eqnarray}
\nonumber &&M_{e}\approx \frac{v_{d}}{6\sqrt{3}}
\begin{pmatrix}
6y_{1}^{e}{\tilde \phi}
&
-6y_{1}^{e}\eta^{2}{\tilde \phi}
&
-6\sqrt{2}y_{1}^{e}\eta{\tilde \phi}
\,\\
-2\sqrt{3}y_{2}^{e}{\tilde \phi}
&
-6\sqrt{3}y_{2}^{e}\eta^{2}{\tilde \phi}
&
-2\sqrt{6}y_{2}^{e}\eta{\tilde \phi}
\,\\
\left[(2\sqrt{2}-\sqrt{6})y_{3}^{e}
+(\sqrt{2}+2\sqrt{6})y_{4}^{e}\right]\eta
&
\sqrt{3}y_{3}^{e}-y_{4}^{e}
&
\left[(1-4\sqrt{3})y_{4}^{e}-(4+\sqrt{3})y_{3}^{e}
\right]\eta^{2}
\end{pmatrix}\,,\\
&&M_{\nu}\approx\frac{v_{u}^{2}}{3\sqrt{6}\Lambda}
\begin{pmatrix}
-\sqrt{6}y_{1}^{\nu}-2y_{3}^{\nu} ~&~   0 ~&~(4\sqrt{3}y_{2}^{\nu}+\sqrt{2}y_{3}^{\nu})\eta   \,\\
0 ~&~ (4\sqrt{3}y_{2}^{\nu}-2\sqrt{2}y_{3}^{\nu})\eta
~&~ -\sqrt{6}y_{1}^{\nu}+y_{3}^{\nu} \,\\
(4\sqrt{3}y_{2}^{\nu}+\sqrt{2}y_{3}^{\nu})\eta
~&~-\sqrt{6}y_{1}^{\nu}+y_{3}^{\nu} ~&~  0
\end{pmatrix}\,.
\end{eqnarray}
Hence the charged lepton masses are approximately
\begin{equation}
m_{e}\approx \frac{2\sqrt{6}}{3}|y^e_{1}|\eta\tilde{\phi}v_d\,,~~~
m_{\mu}\approx\frac{1}{3}|y^e_2|\tilde{\phi}v_d\,,~~~m_{\tau}\approx\frac{1}{6\sqrt{3}}|\sqrt{3}y_{3}^{e}-y_{4}^{e}|v_d\,.
\end{equation}
Notice that our results for charged leptons are consistent with general one presented in table~\ref{tab:A4p-patterns} for the case of $(\bm{r}_{\psi^{C}},\bm{r}_{\psi})=
(\bm{1}^{\prime\prime}\oplus\bm{1}\oplus\bm{1},\bm{3})$, $I=0,  J=K=1$, and the neutrino sector is the case of $\bm{r}_{\nu}=\bm{3}$, $I=J=K=0$ in table~\ref{tab:neutrino-patterns}.

Analogously taking the lowest nontrivial order in each entry of the quark mass matrices, we have
\begin{eqnarray}
\hskip-0.2in &&M_{u}\approx \frac{v_{u}}{6\sqrt{3}}
\begin{pmatrix}
6y_{1}^{u}{\tilde \phi}^{2} ~&~
-6y_{1}^{u}\eta^{2}{\tilde \phi}^{2} ~&~
-6\sqrt{2}y_{1}^{u} \eta{\tilde \phi}^{2} \,\\
\left[(2\sqrt{2}-\sqrt{6})y_{2}^{u}+(\sqrt{2}+2\sqrt{6})y_{3}^{u}\right]\eta{\tilde \phi} ~&~ (\sqrt{3}y_{2}^{u}-y_{3}^{u}){\tilde \phi} ~&~ \left[-(4+\sqrt{3})y_{2}^{u}
+(1-4\sqrt{3})y_{3}^{u}\right]\eta^{2}{\tilde \phi}\,\\
-6\sqrt{3}y_{4}^{u}\eta^{2} ~&~ -2\sqrt{6}y_{4}^{u}\eta ~&~ -2\sqrt{3}y_{4}^{u}
\end{pmatrix}\,,\nonumber\\
\hskip-0.2in &&M_{d}\approx \frac{v_{d}}{3\sqrt{6}}
\begin{pmatrix}
3\sqrt{6}y_{2}^{d}\eta{\tilde\phi} ~&~ (2\sqrt{3}y_{1}^{d}-\sqrt{3}y_{2}^{d})
{\tilde\phi} ~&~ -3\sqrt{3}y_{1}^{d} \eta^{2}{\tilde\phi} \,\\
-3\sqrt{6}y_{1}^{d}\eta^{2}{\tilde\phi} ~&~ -3\sqrt{3}y_{2}^{d}\eta{\tilde\phi}
~&~ -\sqrt{6}(y_{1}^{d}+y_{2}^{d}){\tilde\phi} \,\\
-(2y_{3}^{d}+2\sqrt{3}y_{4}^{d})\eta^{3} ~&~ -2\sqrt{2}(y_{3}^{d}+\sqrt{3}y_{4}^{d})\eta^{2} ~&~ (2y_{3}^{d}+2\sqrt{3}y_{4}^{d})\eta
\end{pmatrix}\,.
\end{eqnarray}
Thus the hierarchical quark masses are approximately given by
\begin{eqnarray}
\nonumber&& m_{u}\approx \frac{1}{\sqrt{3}}|y_{1}^{u}|\tilde{\phi}^{2}v_{u}\,,~~~m_{c}\approx\frac{1}{6\sqrt{3}}|\sqrt{3}y_{2}^{u}-y_{3}^{u}|\,\tilde{\phi}v_{u}\,,~~~m_{t}\approx \frac{1}{3}|y^u_{4}|v_{u}\,,\\
\label{eq:quark-masses-modelA}&& m_{d}\approx \left|\frac{(2y^d_1+5y^d_2)y^d_2}{2y^d_1-y^d_2}\right|\eta^2\tilde{\phi}v_d\,, ~~~m_{s}\approx \frac{1} {3\sqrt{2}}|2y_{1}^{d}-y_{2}^{d}|\tilde{\phi}v_d\,,~~~m_{b}\approx \frac{1}{3}\sqrt{\frac{2}{3}}\,|y_{3}^{d}+\sqrt{3}y_{4}^{d}|\eta v_d\,.~~~~
\end{eqnarray}
Since the quark mass matrices are hierarchical, we can make an estimate for the three quark mixing angles as follows,
\begin{eqnarray}
&&\sin\theta_{12}^{q}\approx  \left|\frac{3\sqrt{2}y_{2}^{d}}
{2y_{1}^{d}-y_{2}^{d}}
-\frac{(2\sqrt{2}-\sqrt{6})y_{2}^{u}
+(\sqrt{2}+2\sqrt{6})y_{3}^{u}}
{\sqrt{3}y_{2}^{u}-y_{3}^{u}}\right|\eta\,,\nonumber \\
&&\sin\theta_{13}^{q}\approx
\left|\frac{4\left[(2-\sqrt{3})y_{2}^{u}
+(1+2\sqrt{3})y_{3}^{u}\right]}
{\sqrt{3}y_{2}^{u}-y_{3}^{u}}\right|\eta^{2}\,,~~~\sin\theta_{23}^{q}\approx
2\sqrt{2}\,\eta\,.
\end{eqnarray}
The above approximations reproduce the numerical values of the quark masses and mixing angles. The up quark sector of this model corresponds to the assignment of
$(\bm{r}_{\psi^{C}},\bm{r}_{\psi})=(\bm{1}^{\prime\prime}
\oplus\bm{1}^{\prime}\oplus\bm{1},\bm{3})$, $I=1,~J=0,~K=2$, the prediction for the up quark mass spectrum in Eq.~\eqref{eq:quark-masses-modelA} are compatible with the general results in table~\ref{tab:A4p-patterns}. The down-type quark sector corresponds to the assignment of $(\bm{r}_{\psi^{C}},\bm{r}_{\psi})=(\bm{\widehat{2}}^{\prime}
\oplus\bm{1},\bm{3})$, $I=J=1,~K=0$. However, in this example, there is an accidental cancellation between the $y^d_3$ and $y^d_4$ terms in the (31) entry of $M_d$ so that its magnitude is of order $\eta^3$ rather than order one at leading order. Thus the approximate estimates in table~\ref{tab:A4p-patterns} are not applicable here and so, although this example explains all hierarchies, one unexplained cancellation is required.

\subsection{Model B}

\begin{figure}[t!]
\begin{center}
\includegraphics[width=0.48\textwidth]{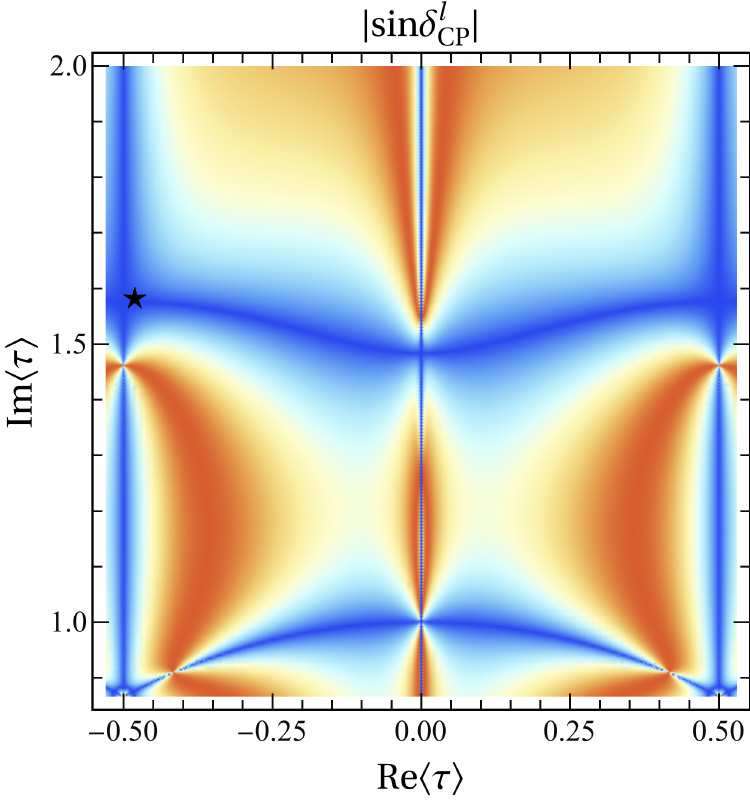}  \includegraphics[width=0.48\textwidth]{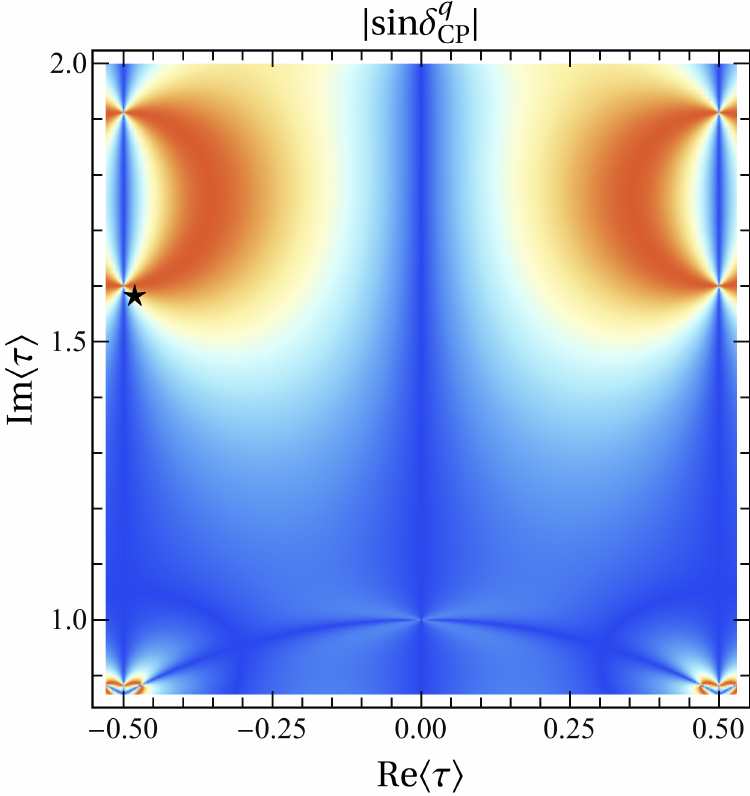}\\
\includegraphics[width=0.48\textwidth]{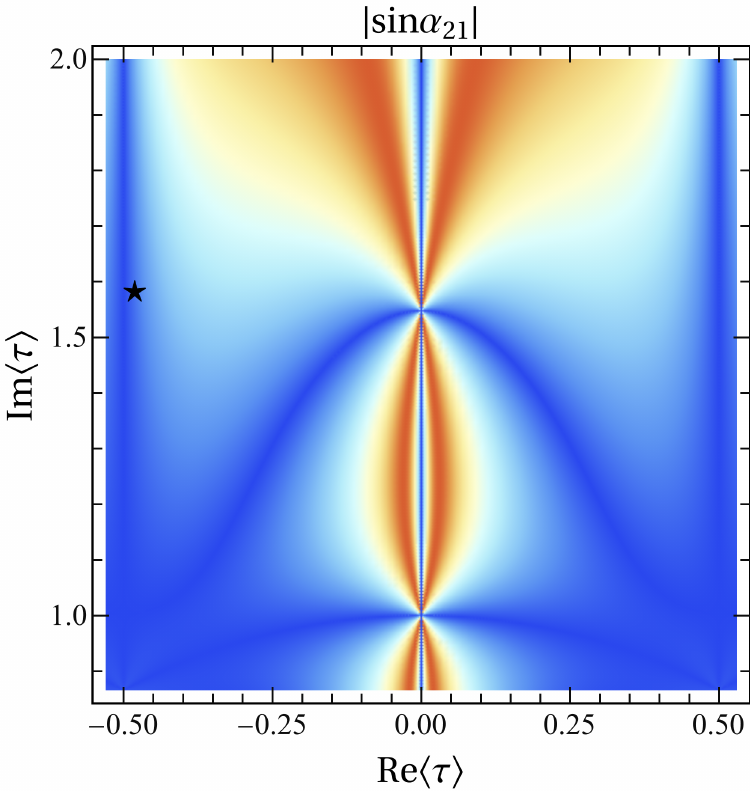} \includegraphics[width=0.48\textwidth]{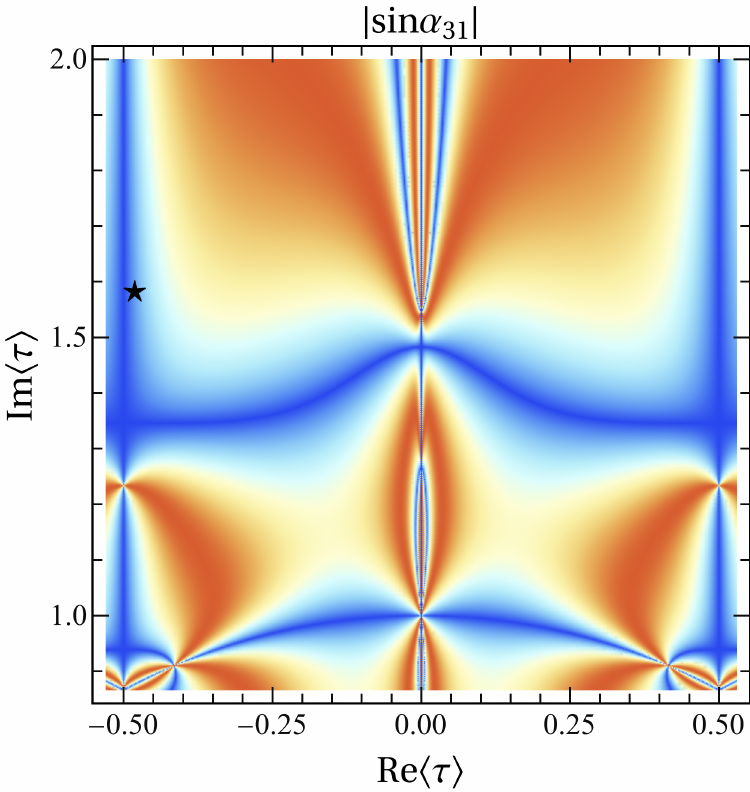} \\
\includegraphics[width=0.8\textwidth]{bar_CaseAB}
\end{center}
\caption{\label{fig:contour-model-B}The contour of $|\sin\delta^{l}_{CP}|$, $|\sin\delta^{q}_{CP}|$, $|\sin\alpha_{21}|$ and $|\sin\alpha_{31}|$ in the plane of $\tau$ for the model B, where black star refers to the best fitting point of $\tau$. The couplings are set to the best fit values in Eq.~\eqref{eq:bf-pars-model-B}}
\end{figure}

The gCP symmetry is also imposed in this model and it constrains all coupling constants to be real. The neutrino masses are generated by the type-I seesaw mechanism. Both left-handed lepton doublets $L$ and right-handed neutrinos are assigned to two triplet $\bm{3}$ of $T'$, while the right-handed charged leptons are singlets of $T'$, i.e.
\begin{eqnarray}\nonumber
&&L\sim (\bm{3},-2)\,,~~e^{c}\sim (\bm{1'},5)\,,~~\mu^{c}\sim (\bm{1''},6)\,,~~\tau^{c}\sim (\bm{1''},8)\,,~~N^{c}\equiv \{N^{c}_{1},N^{c}_{2},N^{c}_{3}\}\sim (\bm{3},2)\,.
\end{eqnarray}
The Higgs fields are invariant with zero modular weight. The modular invariant superpotentials for lepton masses are given by
\begin{eqnarray}\nonumber
\mathcal{W}_{e}&=&y^{e}_{1} {\tilde \phi} e^{c}
(LY^{(4)}_{\bm{3}})_{\bm{1''}}H_{d} + y^{e}_{2} \mu^{c}
(LY^{(4)}_{\bm{3}})_{\bm{1'}}H_{d}+ y^{e}_{3} \tau^{c}
(LY^{(6)}_{\bm{3}A})_{\bm{1'}}H_{d}+ y^{e}_{4} \tau^{c}
(LY^{(6)}_{\bm{3}B})_{\bm{1'}}H_{d}\,,\,\\
\mathcal{W}_{\nu}&=&y^{D}_{1}(LN^{c})_{\bm{1}}H_{u}
+y^{N}_{1}\Lambda  (N^{c}N^{c})_{\bm{1}}Y^{(4)}_{\bm{1}}
+y^{N}_{2}\Lambda\left( (N^{c}N^{c})_{\bm{1''}}Y^{(4)}_{\bm{1'}}\right)_{\bm{1}}+y^{N}_{3}\Lambda \left( (N^{c}N^{c})_{\bm{3}}Y^{(4)}_{\bm{3}}\right)_{\bm{1}}
\,.~~~~~
\end{eqnarray}
The corresponding charged lepton and neutrino mass matrices read as
\begin{eqnarray} \nonumber
&&M_{e}=\frac{v_{d}}{\sqrt{3}}\left(
\begin{array}{ccc}
 y^{e}_{1}\tilde{\phi}\,Y_{\bm{3},3}^{(4)} & y^{e}_{1}\tilde{\phi}\,Y_{\bm{3},2}^{(4)} & y^{e}_{1}\tilde{\phi}\,Y_{\bm{3},1}^{(4)} \,\\
 y^{e}_{2} Y_{\bm{3},2}^{(4)} & y^{e}_{2} Y_{\bm{3},1}^{(4)} & y^{e}_{2} Y_{\bm{3},3}^{(4)} \,\\
 y^{e}_{3} Y_{\bm{3}A,2}^{(6)}+y^{e}_{4} Y_{\bm{3}B,2}^{(6)} & y^{e}_{3} Y_{\bm{3}A,1}^{(6)}+y^{e}_{4} Y_{\bm{3}B,1}^{(6)} & y^{e}_{3} Y_{\bm{3}A,3}^{(6)}+y^{e}_{4} Y_{\bm{3}B,3}^{(6)} \,\\
\end{array}
\right)\,,\,\\
\nonumber&& M_{D}=\frac{y^{D}_{1}v_{u} }{\sqrt{3}}\left(
\begin{array}{ccc}
 1 ~&~ 0 ~&~ 0 \,\\
 0 ~&~ 0 ~&~ 1 \,\\
 0 ~&~ 1 ~&~ 0 \,\\
\end{array}
  \right)\,,\,\\
 &&M_{N}=
 \frac{\Lambda}{3\sqrt{2}}\left(
\begin{array}{ccc}
 2 y^{N}_{3} Y_{\bm{3},1}^{(4)}+\sqrt{6} y^{N}_{1}Y_{\bm{1}}^{(4)}
  &
   -y^{N}_{3} Y_{\bm{3},3}^{(4)}
  &
  \sqrt{6}y^{N}_{2} Y_{\bm{1'}}^{(4)}-y^{N}_{3} Y_{\bm{3},2}^{(4)} \,\\
 -y^{N}_{3} Y_{\bm{3},3}^{(4)}
 &
  \sqrt{6} y^{N}_{2} Y_{\bm{1'}}^{(4)}
  +2y^{N}_{3} Y_{\bm{3},2}^{(4)}
  & \sqrt{6}y^{N}_{1}Y_{\bm{1}}^{(4)}
  -y^{N}_{3} Y_{\bm{3},1}^{(4)} \,\\
 \sqrt{6}y^{N}_{2} Y_{\bm{1'}}^{(4)}
 -y^{N}_{3} Y_{\bm{3},2}^{(4)}
 & \sqrt{6}y^{N}_{1}Y_{\bm{1}}^{(4)}
 -y^{N}_{3} Y_{\bm{3},1}^{(4)}
 &
  2 y^{N}_{3} Y_{\bm{3},3}^{(4)}\,\\
\end{array}
\right)\,.
\end{eqnarray}
The representation assignments and the modular weights of the quark fields are given by,
\begin{eqnarray}\nonumber
&&Q_{L}\sim (\bm{3},k_{Q_{L}})\,,~~u^{c}\sim (\bm{1''},k_{u^{c}})\,,~~c^{c}\sim (\bm{1},k_{c^{c}})\,,~~t^{c}\sim (\bm{1''},k_{t^{c}})\,,\\
&&D_{D}^{c}\equiv \{d^{c},s^{c}\}\sim (\bm{\widehat{2}},k_{D_{D}^{c}})\,,~~b^{c}\sim (\bm{1},,k_{b^{c}})\,,\,
\end{eqnarray}
where $k_{Q_{L}}=-k_{u^{c}}=3-k_{c^{c}}=6-k_{t^{c}}=2-k_{D_{D}^{c}}=6-k_{b^{c}}$.
Consequently the modular invariant superpotentials for the quark Yukawa couplings are of the following form,
\begin{eqnarray}\nonumber
  \mathcal{W}_{u}&=&y^{u}_{1}{\tilde \phi}^{2}u^{c}
  (Q_{L}Y^{(2)}_{\bm{3}})_{\bm{1'}}H_{u} + y^{u}_{2}{\tilde \phi}c^{c}
  (Q_{L}Y^{(4)}_{\bm{3}})_{\bm{1}}H_{u}+ y^{u}_{3}t^{c}
  (Q_{L}Y^{(6)}_{\bm{3}A})_{\bm{1'}}H_{u} + y^{u}_{4}t^{c}
  (Q_{L}Y^{(6)}_{\bm{3}B})_{\bm{1'}}H_{u}\,,\,\\
  \nonumber
  \mathcal{W}_{d}&=&y^{d}_{1}{\tilde \phi}
  \left((D_{D}^{c}Q_{L})_{\bm{\widehat{2}}}
  Y^{(3)}_{\bm{\widehat{2}''}}\right)_{\bm{1}}H_{d}
  +y^{d}_{2}{\tilde \phi}\left((D_{D}^{c}Q_{L})_{\bm{\widehat{2}''}}
  Y^{(3)}_{\bm{\widehat{2}}}\right)_{\bm{1}}H_{d}\,\\
  &&+y^{d}_{3}b^{c}
  (Q_{L}Y^{(6)}_{\bm{3}A})_{\bm{1}}H_{d} + y^{d}_{4}b^{c}
  (Q_{L}Y^{(6)}_{\bm{3}B})_{\bm{1}}H_{d}\,.
\end{eqnarray}
Then we can read off the mass matrices of the up-type quarks and down-type quarks as follow
\begin{eqnarray}\nonumber
    M_{u}&=&\frac{v_{u}}{\sqrt{3}}\left(
\begin{array}{ccc}
y^{u}_{1} \tilde{\phi}^2\, Y_{\bm{3},2}^{(2)} & y^{u}_{1}\tilde{\phi}^2\,Y_{\bm{3},1}^{(2)} & y^{u}_{1}\tilde{\phi}^2\,Y_{\bm{3},3}^{(2)} \,\\
 y^{u}_{2}\tilde{\phi}\, Y_{\bm{3},1}^{(4)} & y^{u}_{2}\tilde{\phi}\, Y_{\bm{3},3}^{(4)} & y^{u}_{2} \tilde{\phi}\,Y_{\bm{3},2}^{(4)} \,\\
 y^{u}_{3} Y_{\bm{3}A,2}^{(6)}+y^{u}_{4} Y_{\bm{3}B,2}^{(6)} & y^{u}_{3} Y_{\bm{3}A,1}^{(6)}+y^{u}_{4} Y_{\bm{3}B,1}^{(6)} & y^{u}_{3} Y_{\bm{3}A,3}^{(6)}+y^{u}_{4} Y_{\bm{3}B,3}^{(6)} \,\\
\end{array}
\right) \,,\,\\
M_{d}&=&\frac{v_{d}}{\sqrt{3}}\left(
\begin{array}{ccc}
\frac{y^{d}_{1}\tilde{\phi}\,Y_{\bm{\widehat{2}''},2}^{(3)}}{\sqrt{2}} & y^{d}_{2}\tilde{\phi}\, Y_{\bm{\widehat{2}},1}^{(3)} & -\frac{y^{d}_{2}\tilde{\phi}\, Y_{\bm{\widehat{2}},2}^{(3)}}{\sqrt{2}}-y^{d}_{1}\tilde{\phi}\,Y_{\bm{\widehat{2}''},1}^{(3)} \,\\
 \frac{y^{d}_{1}\tilde{\phi}\,Y_{\bm{\widehat{2}''},1}^{(3)}}{\sqrt{2}}-y^{d}_{2}\tilde{\phi}\, Y_{\bm{\widehat{2}},2}^{(3)} & y^{d}_{1}\tilde{\phi}\,Y_{\bm{\widehat{2}''},2}^{(3)} & -\frac{y^{d}_{2}\tilde{\phi}\, Y_{\bm{\widehat{2}},1}^{(3)}}{\sqrt{2}} \,\\
 y^{d}_{3} Y_{\bm{3}A,1}^{(6)}+y^{d}_{4} Y_{\bm{3}B,1}^{(6)} & y^{d}_{3} Y_{\bm{3}A,3}^{(6)}+y^{d}_{4} Y_{\bm{3}B,3}^{(6)} & y^{d}_{3} Y_{\bm{3}A,2}^{(6)}+y^{d}_{4} Y_{\bm{3}B,2}^{(6)} \,\\
\end{array}
\right) \,.
\end{eqnarray}
Hence the flavor structure of quarks and leptons are described by totally 18 real parameters including the weighton's VEV $\tilde{\phi}$ in this model. The numerical minimization procedure gives rise to the following point in parameter space which
minimizes the $\chi^2$,
\begin{eqnarray}
\nonumber&&\langle\tau\rangle=-0.4800 + 1.5817i,~~
\langle{\tilde\phi}\rangle=0.001\,,\,\\
&&y^{e}_{1}/y^{e}_{4}=1.4863,
~~y^{e}_{2}/y^{e}_{4}=-1.0218,
~~y^{e}_{3}/y^{e}_{4}=-8.7083,
~~y^{e}_{4}v_{d}=0.7733~\text{GeV}\,,\nonumber\,\\
\nonumber&&y^{N}_{2}/y^{N}_{1}=4.8462,
~~y^{N}_{3}/y^{N}_{1}=-5.1762,
~~(y^{D}_{1}v_{u})^{2}/y^{N}_{1}\Lambda
=67.7914~\text{meV}\,,\,\\
\nonumber&&y^{u}_{1}/y^{u}_{4}=1.3638,
~~y^{u}_{2}/y^{u}_{4}=1.7495,
~~y^{u}_{3}/y^{u}_{4}=0.5687,
~~y^{u}_{4}v_{u}=837.321~\text{GeV}\,,\,\\
\label{eq:bf-pars-model-B}&&y^{d}_{1}/y^{d}_{3}=9.2752,
~~y^{d}_{2}/y^{d}_{3}=-1.4005,
~~y^{d}_{4}/y^{d}_{3}=-0.7073,
~~y^{d}_{3}v_{d}=3.6926~\text{GeV}\,,
\end{eqnarray}
with $\chi^{2}_{\text{min}}=12.5$. We see that the VEV of $\tau$ is very close to the left vertical boundary of the fundamental domain. The masses and mixing parameters of leptons and quarks at the above best fit point are determined to be
\begin{eqnarray}
\nonumber&& \sin^{2}\theta^{l}_{12}= 0.323\,,\quad \sin^{2}\theta^{l}_{13}=0.02212\,,\quad \sin^{2}\theta^{l}_{23}=0.479\,,\quad \delta^{l}_{CP}=180.5^{\circ}\,,\,\\
\nonumber&& \alpha_{21}=1.022 \pi\,,\quad \alpha_{31}=1.047\pi\,,\quad m_e/m_{\mu}=0.004786\,,\quad m_{\mu}/m_{\tau}=0.05908\,,\,\\
\nonumber&& m_1=13.50~\text{meV}\,,\quad m_2=16.03~\text{meV}\,,\quad m_3=51.94~\text{meV}\,,\quad m_{\beta\beta}=2.80~\text{meV}\,,\,\\
\nonumber&& \theta^q_{12}=0.2290\,,\quad \theta^q_{13}=0.002836\,,\quad \theta^q_{23}=0.04022\,,\quad \delta^q_{CP}=43.30^{\circ} \,,\,\\
&& m_u / m_c=0.003025\,,\quad m_c / m_t=0.002546\,,\quad m_d/m_s=0.06120\,,\quad m_s / m_b=0.01926\,.
\end{eqnarray}
All these observables are compatible with the experimental data at $3\sigma$ levels. The leptonic Dirac and Majorana CP violation phases are predicted to be very close to $\pi$, and the effective Majorana neutrino mass $m_{\beta\beta}$ is too small to be detectable. Let $\tau=-0.5+1.5817i$ and the coupling constants at the best fit values in Eq.~\eqref{eq:bf-pars-model-B}, then all CP phases would be conserved and the flavor observables are determined to be
\begin{eqnarray}
\nonumber&& \sin^{2}\theta^{l}_{12}=0.322\,,\quad \sin^{2}\theta^{l}_{13}=0.02147\,,\quad \sin^{2}\theta^{l}_{23}=0.479\,,\quad \delta^{l}_{CP}=180^{\circ}\,,\\
\nonumber&& \alpha_{21}=\pi\,,\quad \alpha_{31}=\pi\,,\quad m_e/m_{\mu}=0.004788\,,\quad m_{\mu}/m_{\tau}=0.05908\,,\\
\nonumber&&  m_1=13.61~\text{meV}\,,\quad m_2=16.12~\text{meV}\,,\quad m_3=52.40~\text{meV}\,,\\
\nonumber&& m_{\beta\beta}=2.81~\text{meV}\,,\quad \theta^q_{12}=0.2287\,,\quad \theta^q_{13}=0.001926\,,\quad \theta^q_{23}=0.04025\,,\quad \delta^q_{CP}=0^{\circ} \,,\\
 && m_u/m_c=0.003025\,,\quad m_c / m_t=0.002546\,,\quad m_d/m_s=0.06120\,,\quad m_s / m_b=0.01926\,.
\end{eqnarray}
Therefore the small departure of $\text{Re}(\tau)$ from $-1/2$ is the unique of CP violation and its influence on fermion masses and mixing angles is so small that it can be ignored. We show the contour plots of $|\sin\delta^{l}_{CP}|$, $|\sin\delta^{q}_{CP}|$, $|\sin\alpha_{21}|$ and $|\sin\alpha_{31}|$ in the complex $\tau$  plane in figure~\ref{fig:contour-model-B}. One sees that the CP violation phases are very sensitive to $\tau$.

It is interesting to perform an analytical approximation for the lepton and quark mass matrices. The leading order approximation for each element of the lepton mass matrices gives,
\begin{eqnarray}
\nonumber M_{e}&\approx& \frac{v_{d}}{6\sqrt{3}}
\begin{pmatrix}
-6\sqrt{3}y_{1}^{e}\omega^{2}\eta^{2}\tilde{\phi}
~&~ 2\sqrt{6}y_{1}^{e} \omega\eta\tilde{\phi}
~&~-2\sqrt{3}y_{1}^{e}\tilde{\phi} \,\\
2\sqrt{6}y_{2}^{e}\omega\eta ~&~ -2\sqrt{3}y_{2}^{e}
~&~  -6\sqrt{3}y_{2}^{e}\omega^{2}\eta^{2} \,\\
[(\sqrt{6}-2\sqrt{2})y_{3}^{e}-(\sqrt{2}+2\sqrt{6})y_{4}^{e}]  \omega\eta      ~&~\sqrt{3}y_{3}^{e}-y_{4}^{e}    ~&~ -[(4+\sqrt{3})y_{3}^{e}+(4\sqrt{3}-1)y_{4}^{e}]\omega^{2}\eta^{2}
\end{pmatrix}\,,\\
\nonumber\\
M_{D}&=&\frac{y^{D}_{1}v_{u} }{\sqrt{3}}\left(
\begin{array}{ccc}
 1 ~&~ 0 ~&~ 0 \,\\
 0 ~&~ 0 ~&~ 1 \,\\
 0 ~&~ 1 ~&~ 0 \,\\
\end{array}
  \right)\,,\,\nonumber\\
M_{N}&=&
 \frac{\Lambda}{3\sqrt{6}}\left(
\begin{array}{ccc}
-\sqrt{6}y^{N}_{1}-2y^{N}_{3} & 3y^{N}_{3}\omega^{2}\eta^{2}  &
-(4\sqrt{3}y^{N}_{2}+\sqrt{2}y^{N}_{3})\omega\eta \,\\
3y^{N}_{3}\omega^{2}\eta^{2} &(-4\sqrt{3}y^{N}_{2}+2\sqrt{2}y^{N}_{3})\omega\eta
& -\sqrt{6}y^{N}_{1}+y^{N}_{3} \, \\
-(4\sqrt{3}y^{N}_{2} +\sqrt{2}y^{N}_{3})\omega\eta
&-\sqrt{6}y^{N}_{1} +y^{N}_{3}
&-6y^{N}_{3}\omega^{2}\eta^{2}\,\\
\end{array}
\right) \,,
\end{eqnarray}
where $\eta\equiv3\sqrt{2}|q^{1/3}|\approx0.155$ for the best fit value of $\tau$.  Hence  the charged lepton mass hierarchy can be naturally explained,
\begin{eqnarray}
m_e=\frac{1}{3}|y^e_{1}|\tilde{\phi}v_d\,,~~~
m_{\mu}=\frac{2\sqrt{2}}{3}\left|\frac{(1-\sqrt{3})y^e_3+(1+\sqrt{3})y^e_4}{\sqrt{3}y^e_3-y^e_4}y^e_2\right|\eta v_d\,,~~~m_{\tau}=\frac{1}{6\sqrt{3}}|\sqrt{3}y_{3}^{e}-y_{4}^{e}|v_{d}\,.
\end{eqnarray}
This is compatible with the general results in table~\ref{tab:A4p-patterns} for the case of $(\bm{r}_{\psi^{C}},\bm{r}_{\psi})=
(\bm{1}^{\prime\prime}\oplus\bm{1}^{\prime\prime}\oplus\bm{1}^{\prime},\bm{3})$, $I=J=0,~K=1$.

Similarly taking the lowest non-trivial order in each entry of the quark mass matrices $M_u$ and $M_d$, we find
\begin{small}
\begin{eqnarray}
&&\hskip-0.2in M_{u}\approx \frac{v_{u}}{3\sqrt{3}}
\begin{pmatrix}
3\sqrt{2}y_{1}^{u} \omega\eta\tilde{\phi}^{2} ~&~ 3y_{1}^{u}\tilde{\phi}^{2}   ~&~ -3y_{1}^{u}\omega^{2}\eta^{2}\tilde{\phi}^{2}\, \\
-\sqrt{3}\,y_{2}^{u}{\tilde\phi} ~&~ -3\sqrt{3}\,y_{2}^{u} \omega^{2}\eta^{2}{\tilde\phi}  ~&~ \sqrt{6}\,y_{2}^{u}\omega\eta{\tilde\phi} \, \\
-(\sqrt{2}\,y_{3}^{u}+\sqrt{6}\,y_{4}^{u})\omega\eta
~&~ (\sqrt{2}\,y_{3}^{u}+\sqrt{6}\,y_{4}^{u})\eta^{3} ~&~
-(2y_{3}^{u}+2\sqrt{3}\,y_{4}^{u})\omega^{2}\eta^{2}
  \end{pmatrix}\,, \nonumber\\
&&\hskip-0.2in M_{d}\approx \frac{v_{d}}{6\sqrt{3}}
\begin{pmatrix}
-3\sqrt{6}y_{1}^{d}\omega^{2}\eta^{2}{\tilde\phi}
~&~ -6\sqrt{3}y_{2}^{d}\omega\eta{\tilde\phi}
~&~ (2\sqrt{6}y_{1}^{d}-\sqrt{6}y_{2}^{d}) {\tilde\phi} \,\\
-2\sqrt{3}(y_{1}^{d}+y_{2}^{d}){\tilde\phi}
~&~ -6\sqrt{3}y_{1}^{d}\omega^{2}\eta^{2}{\tilde\phi}
~&~3\sqrt{6}y_{2}^{d} \omega\eta{\tilde\phi} \,\\
\sqrt{3}y_{3}^{d}-y_{4}^{d}
 ~&~ [(1-4\sqrt{3})y_{4}^{d}-(4+\sqrt{3})y_{3}^{d}]\omega^{2}\eta^{2}
~&~-[(2\sqrt{2}-\sqrt{6})y_{3}^{d}+(\sqrt{2}+2\sqrt{6})y_{4}^{d}] \omega\eta
\end{pmatrix}\,.~~~~~~
\end{eqnarray}
\end{small}
Thus we can obtain an approximate estimation of the quark masses as follow,
\begin{eqnarray}
&&m_{u}\approx\frac{1}{\sqrt{3}}|y_{1}^{u}|\tilde{\phi}^{2}v_{u}\,,
~~~m_{c}\approx\frac{2\sqrt{2}}{3}|y_{2}^{u}|\eta\tilde{\phi}v_{u}\,, ~~~m_{t}\approx\frac{1}{3}\sqrt{\frac{2}{3}}\,|y_{3}^{u}
  +\sqrt{3}y_{4}^{u}|\eta v_{u}\,,\nonumber\\
\nonumber&&m_{d}\approx\left|\frac{4y_{1}^{d}+y_{2}^{d}}{3}
+\frac{4(y_{1}^{d}+y_{2}^{d})(y_{3}^{d}+\sqrt{3}y_{4}^{d})}{3(\sqrt{3}y_{3}^{d}-y_{4}^{d})}\right|\eta^{2}\tilde{\phi}v_{d}\,,\\
&&m_{s}\approx\frac{1}{3\sqrt{2}}\left|2y_{1}^{d}-y_{2}^{d}\right|\tilde{\phi}v_{d}\,,
~~~m_{b}\approx\frac{1}{6\sqrt{3}}|\sqrt{3}y_{3}^{d}-y_{4}^{d}|v_{d}\,. \label{eq:quark-masses-modelB}
\end{eqnarray}
One can also make an estimate for the three quark mixing angles,
\begin{eqnarray}
&&\sin\theta_{12}^{q}\approx2\sqrt{2}
\left|\frac{y_{1}^{d}+y_{2}^{d}}{2y_{1}^{d}-y_{2}^{d}}\right|\eta\,,~~~\sin\theta_{23}^{q}\approx2\sqrt{2}\left|\frac{y_{3}^{d}+\sqrt{3}y_{4}^{d}}{\sqrt{3}y_{3}^{d}-y_{4}^{d}}\right|\eta\,,
~~~\sin\theta_{13}^{q}\approx{\cal O}(\eta^{2})\,. \label{eq:quark-angles-modelB}
\end{eqnarray}
The approximate formulae of $\sin\theta^q_{13}$ is too lengthy to provide some useful insight, consequently we don't present its cumbersome expression here. The above approximations well reproduce the numerical values of the quark masses and mixing angles. It is easy to find that our results are consistent with the general results in table~\ref{tab:A4p-patterns} for the case of $(\bm{r}_{\psi^{C}},\bm{r}_{\psi})=(\bm{\widehat{2}}\oplus\bm{1},\bm{3})$, $I=J=1,~K=0$ in the down quark sector. Regarding the up quark sector, the coefficient of the order one term in the $(32)$ entry of $M_{u}$ is very small because of the accidental cancellation between the $y^u_3$ and $y^u_4$ terms, thus the predictions for the up quark mass pattern in Eq.~\eqref{eq:quark-masses-modelB} and mixing angles in Eq.~\eqref{eq:quark-angles-modelB} are different from the general estimates in table~\ref{tab:A4p-patterns}. As in the previous example, although this example explains all hierarchies, one unexplained cancellation is required.

\section{\label{sec:conclusion}Conclusion}

In this paper we have systematically developed the weighton mechanism for natural quark and charged lepton mass hierarchies in the framework of modular symmetry with a single modulus field. The weighton $\phi$, which is defined as a complete singlet with unit modular weight, leads to fermion mass suppression by powers of $\tilde{\phi}$, which is
the VEV of the field scaled by a flavour cut-off, while further mixing angle suppression comes from powers of the small parameter, $q=e^{i2\pi \tau}$. Assuming some fields transform as triplets under the finite modular symmetry, with general assignments for the other fields, we have performed a complete analysis for the levels $N=3, 4, 5$, expressing fermion masses and mixings in terms of powers of the small parameters $\tilde{\phi}$ and $q$. We have also shown that the problem of CP violation can be decoupled from the mass and mixings if one works near to the CP boundary. We have presented two examples in detail, based on the modular group $T'$, close to the CP boundary of $\tau$, which can naturally explain mass hierarchies, mixing parameters and CP violation for quarks and leptons.

It is worth recapping that the success of this approach relies in part on the $q$-expansion, which is due to the intrinsic properties of modular symmetry, as well as CP symmetry as follows.
Modular symmetry in combination with supersymmetry plays a crucial role in constraining the Yukawa couplings to be holomorphic modular forms of level $N$ which can be written as convergent infinite series of $q\equiv e^{2\pi i\tau}$. Every point of $\tau$ in the upper half complex plane can be mapped into the fundamental domain $\mathcal{D}=\left\{\tau\,\big|\,\text{Im}(\tau)>0, |\text{Re}(\tau)|\leq \frac{1}{2}, |\tau|\geq1\right\}$ by a modular transformation, but no two points in the interior of $\mathcal{D}$ are related under modular group. Without loss of generality, one can limit the modulus $\tau$ in the fundamental domain $\mathcal{D}$. Thus $|q|\leq e^{-\sqrt{3}\pi}\simeq0.0043\ll1$ is quite small and the first nontrivial term in the $q$-expansion of a modular form gives a leading approximation to its order of magnitude.  The CP fixed points of $\tau$ are on the imaginary axis and at the boundary of the fundamental domain, and the theory still enjoys a residual CP symmetry if the value of $\tau$ is in the vicinity of these CP fixed points. The Yukawa couplings are also subject to constraint from this residual CP symmetry, and a small deviation from the CP fixed points can lead to large CP violation.
The Yukawa couplings of SM fermions have specific transformation properties under the action of the modular group and gCP symmetry, as shown in Eq.~\eqref{eq:me-mnu-gamma} and Eq.~\eqref{eq:me-mnu-CP-modsym}. By considering the $T$ transformation in the basis where the $T$ generator is represented by a diagonal matrix, one can determine the order of magnitude of the Yukawa couplings which can be parameterized as $|q|^r$, where the power $r$ depends on the transformation of the fermion fields under the action of $T$.

We have seen that the other necessary ingredient for explaining fermion mass hierarchies is the weighton mechanism, which is the main subject of this paper.
The modular weights of the fermion fields can play the role of Froggatt-Nielsen charges, and an additional SM and finite modular symmetry singlet field $\phi$ with non-zero modular weight called weighton plays the role of the FN flavon. The presence of the weighton $\phi$ leads to additional suppression of Yukawa couplings by $\tilde{\phi}^k$, where $\tilde{\phi}\equiv\langle\phi\rangle/M_{fl}$ is a small ratio and $M_{fl}$ denotes the cut-off flavour scale. The power $k$ depends on the modular weights of fermion fields. We have performed a comprehensive analysis of the hierarchical patterns of fermion masses and rotation angles for the levels $N=3, 4, 5$, when at least either the electroweak left-handed fields or the charge conjugation of right-handed fields transform as a triplet of the finite modular group $\Gamma'_N$, and the general results are collected in Appendix~\ref{app:patterns-masses}.

As already remarked, the results in Appendix~\ref{app:patterns-masses} are no longer valid in certain models if there exists accidental cancellation between different contributions, and we have seen examples of this in Models A and B, which are two viable models based on the modular symmetry $\Gamma'_3\cong T'$. Assuming one accidental cancellation, the masses and mixing of quarks and leptons can be well accommodated for a common value of $\tau$ in both models. The fermion mass and mixing hierarchies emerge from the synergy between modular forms and weighton. In the first model, the neutrino masses are described by the Weinberg operator, and the VEV of $\tau$ is near the imaginary axis. In the residual CP symmetric limit for which $\tau$ is exactly on the imaginary axis and the values of couplings remain unchanged, the predictions for fermion masses and mixing angles only change a bit and they are still compatible with the experimental data although all the CP violation phases are conserved. The small real part of $\tau$ is responsible for the CP violation in both quark and lepton sectors. The VEV of $\tau$ is close to the left vertical boundary of the fundamental domain in the second model. Likewise the small departure from CP symmetric limit $\text{Re}(\tau)=-1/2$ introduces very small corrections to masses and mixing angles, while it generates the non-trivial CP phases of quarks and leptons.

Finally we remark that, in the non-holomorphic cousin of the modular flavor symmetry~\cite{Ding:2020zxw,Qu:2024rns}, the Yukawa couplings are polyharmonic Maa{\ss} forms of level $N$, and one can estimate the order of magnitude of the Yukawa couplings in a similar manner. Since the polyharmonic Maa{\ss} forms have non-holomorphic part for the weight $k\leq2$, the magnitudes of the Yukawa couplings could be different from these predictions in holomorphic modular symmetry. Thus one expects to obtain new hierarchical patterns of fermion masses and mixing.

In conclusion, understanding the flavour puzzle in the framework of modular symmetry, in which the three fermion families are embedded into at least one triplet representation of the finite modular symmetry, seems to require at least one weighton field which is responsible for the fermion mass hierarchies. The combination of the $q$-expansion and the weighton mechanism provides a natural framework for understanding all the hierarchies in the quark and lepton sectors.
Hopefully the general results presented in paper will provide a useful resource in this endeavour.

\section*{Acknowledgements}

GJD is supported by the National Natural Science Foundation of China under Grant No.~12375104. JNL is supported by the National Natural Science Foundation of China with Grant No.~12147110 and the China Post-doctoral Science Foundation under Grant No.~2021M70. MHW is supported by the National Natural Science Foundation of China under Grant No. 12275263. SFK acknowledges CERN hospitality, the STFC Consolidated Grant ST/L000296/1 and the European Union's Horizon 2020 Research and Innovation programme under Marie Sk\l{}odowska-Curie grant agreement HIDDeN European ITN project (H2020-MSCA-ITN-2019//860881-HIDDeN).

\begin{appendix}

\section{\label{app:Tp-group}Modular group $\Gamma'_3\cong T^{\prime}$ and modular forms of level $N=3$}

The homogeneous finite modular group $\Gamma'_3$ of level $N=3$ is isomorphic to the $T'$ group which is the double covering of the tetrahedral group $A_4$. All the elements of $\Gamma'_3\cong T^{\prime}$ can be generated by the modular generators $S$ and $T$ satisfying the following relations
\begin{equation}
S^{4}=(ST)^{3}=T^{3}=1,~~S^2T = TS^2\,.
\end{equation}
Since $S^2$ commutes with all group elements, the center of $T^{\prime}$ is $Z^{S^2}_2\equiv\left\{1, S^2\right\}$. Notice that $A_4$ is not a subgroup of $T'$, although the quotient group $T'/Z^{S^2}_2$ is isomorphic to $A_4$. The finite modular group $\Gamma'_3\cong T^{\prime}$ has three singlet representations $\bm{1}$, $\bm{1'}$, $\bm{1''}$, three doublet representations $\bm{\widehat{2}}$, $\bm{\widehat{2}'}$, $\bm{\widehat{2}''}$, and a triplet representation $\bm{3}$. In the odd-dimensional representations $\bm{1}$, $\bm{1'}$, $\bm{1''}$ and $\bm{3}$, we have $S^2=\mathbb{1}$ so that $T'$ and $A_4$ are represented by the same set of matrices. In the doublet representations $\bm{\widehat{2}}$, $\bm{\widehat{2}'}$, $\bm{\widehat{2}''}$ which are faithful representations of $T'$, we have $S^2=-\mathbb{1}$. In this work, we adopt the same basis as Ref.~\cite{Ding:2022aoe}, The representation matrices of the generators $S$ and $T$ are listed in table~\ref{tab:representation-matrices-Tp}.

\begin{table}[t!]
\centering
\begin{tabular}{|c|c|c|}
\hline  \hline

~&~  $S$ ~&~ $T$ \,\\  \hline

$\bm{1}$ ~&~ $1$ ~&~ $1$  \,\\  \hline

$\bm{1'}$ ~&~  $1$ ~&~ $\omega$  \,\\  \hline

$\bm{1''}$ ~&~ $1$ ~&~ $\omega^{2}$ \,\\  \hline

$\bm{\widehat{2}}$ ~&~ $-\frac{i}{\sqrt{3}}
\begin{pmatrix}
1  ~&  \sqrt{2}  \,\\
\sqrt{2}  ~&  -1
\end{pmatrix}$ ~&~ $\begin{pmatrix}
\omega & 0 \,\\
0 & 1 \,\\
\end{pmatrix}$  \,\\  \hline

$\bm{\widehat{2}'}$ ~&~ $-\frac{i}{\sqrt{3}}
\begin{pmatrix}
1  ~&  \sqrt{2}  \,\\
\sqrt{2}  ~&  -1
\end{pmatrix}$ ~&~ $\begin{pmatrix}
\omega^{2} & 0 \,\\
0 & \omega \,\\
\end{pmatrix}$ \,\\  \hline

$\bm{\widehat{2}''}$ ~&~ $-\frac{i}{\sqrt{3}}
\begin{pmatrix}
1  ~&  \sqrt{2}  \,\\
\sqrt{2}  ~&  -1
\end{pmatrix}$ ~&~ $\begin{pmatrix}
1 & 0 \,\\
0 & \omega^{2} \,\\
\end{pmatrix}$ \,\\  \hline

$\bm{3}$ ~&~ $\frac{1}{3}\begin{pmatrix}
-1~& 2 &~2\,\\
2~& -1 &~2\,\\
2~& 2 &~-1
\end{pmatrix}$  ~&~ $\begin{pmatrix}
1~& 0~& 0\,\\
0~& \omega~& 0\,\\
0~& 0~& \omega^2
\end{pmatrix} $ \,\\  \hline \hline
\end{tabular}
\caption{\label{tab:representation-matrices-Tp} The representation matrices of the generators $S$ and $T$ in different irreducible representations of $\Gamma'_3\cong T'$. Here $\omega=e^{2\pi i/3}=-\frac{1}{2}+i\frac{\sqrt{3}}{2}$ is the cube root of unit.}
\end{table}

We proceed to report the Clebsch-Gordan coefficients in our basis, they are necessary in the construction of $T'$ invariant model. The tensor products of two $T'$ doublets decompose into a singlet and a triplet as follow,
\begin{eqnarray}
\nonumber\begin{pmatrix}
a_1 \\
a_2
\end{pmatrix}_{\bm{\widehat{2}}\,(\bm{\widehat{2}'})}\otimes \begin{pmatrix}
b_1 \\
b_2
\end{pmatrix}_{\bm{\widehat{2}}\, (\bm{\widehat{2}''})}&=&\Big(\frac{1}{\sqrt{2}}a_1b_2-\frac{1}{\sqrt{2}}a_2b_1\Big)_{\bm{1'}} \oplus \left(\begin{array}{c} a_2b_2 \,\\
\frac{1}{\sqrt{2}}(a_1b_2+a_2b_1)  \,\\
-a_1b_1 \end{array}\right)_{\bm{3}}\,,\\
\nonumber\begin{pmatrix}
a_1 \\
a_2
\end{pmatrix}_{\bm{\widehat{2}'}\,(\bm{\widehat{2}})}\otimes \begin{pmatrix}
b_1 \\
b_2
\end{pmatrix}_{\bm{\widehat{2}'}\, (\bm{\widehat{2}''})}&=&\Big(\frac{1}{\sqrt{2}}a_1b_2-\frac{1}{\sqrt{2}}a_2b_1\Big)_{\bm{1'}} \oplus \left(\begin{array}{c}  \frac{1}{\sqrt{2}}(a_1b_2+a_2b_1)\,\\
-a_1b_1\,\\
a_2b_2  \end{array}\right)_{\bm{3}}\,,\\
\begin{pmatrix}
a_1 \\
a_2
\end{pmatrix}_{\bm{\widehat{2}''}\,(\bm{\widehat{2}})}\otimes \begin{pmatrix}
b_1 \\
b_2
\end{pmatrix}_{\bm{\widehat{2}''}\, (\bm{\widehat{2}'})}&=&\Big(\frac{1}{\sqrt{2}}a_1b_2-\frac{1}{\sqrt{2}}a_2b_1\Big)_{\bm{1''}} \oplus \left(\begin{array}{c} -a_1b_1\,\\
a_2b_2 \,\\
\frac{1}{\sqrt{2}}(a_1b_2+a_2b_1) \end{array}\right)_{\bm{3}}\,.
\end{eqnarray}
The multiplication laws for the contraction of doublets and triplet are given by
\begin{eqnarray}
\nonumber\begin{pmatrix}
a_1 \\
a_2
\end{pmatrix}_{\bm{\widehat{2}}}\otimes \begin{pmatrix}
b_1 \\
b_2 \\
b_3
\end{pmatrix}_{\bm{3}}&=&\left(\begin{array}{c} \frac{1}{\sqrt{3}}a_1b_1+\sqrt{\frac{2}{3}}a_2b_2 \,\\
 -\frac{1}{\sqrt{3}}a_2b_1+\sqrt{\frac{2}{3}}a_1b_3 \end{array}\right)_{\bm{\widehat{2}}} \oplus\left(\begin{array}{c} \frac{1}{\sqrt{3}}a_1b_2+\sqrt{\frac{2}{3}}a_2b_3 \,\\
-\frac{1}{\sqrt{3}}a_2b_2+\sqrt{\frac{2}{3}}a_1b_1  \end{array}\right)_{\bm{\widehat{2}'}}\\
\nonumber&&\quad \oplus \left(\begin{array}{c} \frac{1}{\sqrt{3}}a_1b_3+\sqrt{\frac{2}{3}}a_2b_1 \,\\
-\frac{1}{\sqrt{3}}a_2b_3+\sqrt{\frac{2}{3}}a_1b_2  \end{array}\right)_{\bm{\widehat{2}''}} \,,\\
\nonumber\begin{pmatrix}
a_1 \\
a_2
\end{pmatrix}_{\bm{\widehat{2}'}}\otimes \begin{pmatrix}
b_1 \\
b_2 \\
b_3
\end{pmatrix}_{\bm{3}}&=&\left(\begin{array}{c} \frac{1}{\sqrt{3}}a_1b_3+\sqrt{\frac{2}{3}}a_2b_1 \,\\
-\frac{1}{\sqrt{3}}a_2b_3+\sqrt{\frac{2}{3}}a_1b_2  \end{array}\right)_{\bm{\widehat{2}}} \oplus \left(\begin{array}{c} \frac{1}{\sqrt{3}}a_1b_1+\sqrt{\frac{2}{3}}a_2b_2 \,\\
 -\frac{1}{\sqrt{3}}a_2b_1+\sqrt{\frac{2}{3}}a_1b_3 \end{array}\right)_{\bm{\widehat{2}'}}\\
\nonumber&&\quad\oplus \left(\begin{array}{c}
\frac{1}{\sqrt{3}}a_1b_2+\sqrt{\frac{2}{3}}a_2b_3 \,\\
-\frac{1}{\sqrt{3}}a_2b_2+\sqrt{\frac{2}{3}}a_1b_1  \end{array}\right)_{\bm{\widehat{2}''}} \,,\\
\nonumber\begin{pmatrix}
a_1 \\
a_2
\end{pmatrix}_{\bm{\widehat{2}''}}\otimes \begin{pmatrix}
b_1 \\
b_2 \\
b_3
\end{pmatrix}_{\bm{3}}&=&\left(\begin{array}{c} \frac{1}{\sqrt{3}}a_1b_2+\sqrt{\frac{2}{3}}a_2b_3 \,\\
-\frac{1}{\sqrt{3}}a_2b_2+\sqrt{\frac{2}{3}}a_1b_1  \end{array}\right)_{\bm{\widehat{2}}} \oplus
\left(\begin{array}{c} \frac{1}{\sqrt{3}}a_1b_3+\sqrt{\frac{2}{3}}a_2b_1 \,\\
-\frac{1}{\sqrt{3}}a_2b_3+\sqrt{\frac{2}{3}}a_1b_2  \end{array}\right)_{\bm{\widehat{2}'}} \\
&&\quad \oplus
\left(\begin{array}{c} \frac{1}{\sqrt{3}}a_1b_1+\sqrt{\frac{2}{3}}a_2b_2 \,\\
 -\frac{1}{\sqrt{3}}a_2b_1+\sqrt{\frac{2}{3}}a_1b_3 \end{array}\right)_{\bm{\widehat{2}''}}\,.
\end{eqnarray}
Finally the contraction rules of two triplets are given by
\begin{eqnarray}
\nonumber\begin{pmatrix}
a_1 \\
a_2 \\
a_3
\end{pmatrix}_{\mathbf{3}}\otimes \begin{pmatrix}
b_1 \\
b_2\\
b_3
\end{pmatrix}_{\mathbf{3}}&=& \Big(\frac{1}{\sqrt{3}} a_1b_1 + \frac{1}{\sqrt{3}}a_2b_3 + \frac{1}{\sqrt{3}}a_3b_2 \Big)_{\bm{1}}\oplus\Big(\frac{1}{\sqrt{3}}a_3b_3 +\frac{1}{\sqrt{3}}a_1b_2 + \frac{1}{\sqrt{3}}a_2b_1 \Big)_{\bm{1'}}\\
\nonumber&&\oplus\left(\frac{1}{\sqrt{3}}a_2b_2 + \frac{1}{\sqrt{3}} a_1b_3 +\frac{1}{\sqrt{3}} a_3b_1 \right)_{\bm{1''}} \oplus
\left(\begin{array}{c} \frac{2}{\sqrt{6}}a_1b_1 - \frac{1}{\sqrt{6}}a_2b_3 -\frac{1}{\sqrt{6}} a_3b_2 \,\\
\frac{2}{\sqrt{6}}a_3b_3 - \frac{1}{\sqrt{6}}a_1b_2 - \frac{1}{\sqrt{6}}a_2b_1  \,\\
\frac{2}{\sqrt{6}}a_2b_2 - \frac{1}{\sqrt{6}}a_1b_3 - \frac{1}{\sqrt{6}}a_3b_1 \end{array}\right)_{\bm{3}_S} \\
&&\oplus\left(\begin{array}{c} \frac{1}{\sqrt{2}}a_2b_3 - \frac{1}{\sqrt{2}}a_3b_2 \,\\
\frac{1}{\sqrt{2}}a_1b_2 - \frac{1}{\sqrt{2}}a_2b_1  \,\\
\frac{1}{\sqrt{2}}a_3b_1 - \frac{1}{\sqrt{2}}a_1b_3 \end{array}\right)_{\bm{3}_A}\,.
\end{eqnarray}

\subsection{\label{subsec:modular-forms-level-3}Modular forms of level $N=3$ }

There are $k+1$ linearly independent modular forms at level $N=3$ and weight $k$, and they can be expressed as linear combination of $\left(\frac{\eta^{3}(\tau /3 )}{\eta(\tau)}\right)^k$, $\frac{\eta^{3}(3\tau)}{\eta(\tau)}\left(\frac{\eta^{3}(\tau /3 )}{\eta(\tau)}\right)^{k-1}$, $\left(\frac{\eta^{3}(3\tau)}{\eta(\tau)}\right)^2\left(\frac{\eta^{3}(\tau /3 )}{\eta(\tau)}\right)^{k-2}$,\ldots, $\left(\frac{\eta^{3}(3\tau)}{\eta(\tau)}\right)^k$~\cite{Liu:2019khw,Lu:2019vgm,Ding:2022aoe}, where $\eta(\tau)$ is the Dedekind eta function defined as,
\begin{equation}
\eta(\tau)=q^{1/24}\prod^{+\infty}_{n=1}(1-q^n)\,,~~~q=e^{2\pi i\tau}\,.
\end{equation}
The two modular forms of weight 1 and level 3 can be arranged into a modular multiplet $Y^{(1)}_{\bm{\widehat{2}}}(\tau)$ transforming as a doublet $\bm{\widehat{2}}$ of $\Gamma'_3 \cong T'$ up to the automorphic factor~\cite{Liu:2019khw,Lu:2019vgm,Ding:2022aoe}:
\begin{equation}
\label{eq:modular_space}
Y^{(1)}_{\bm{\widehat{2}}}(\tau)=\begin{pmatrix}
Y_1(\tau) \,\\
Y_2(\tau)
\end{pmatrix}\,,
\end{equation}
with
\begin{eqnarray}
\nonumber Y_1(\tau)&=&\sqrt{2}\,\frac{\eta^{3}(3\tau)}{\eta(\tau)}=3\sqrt{2}\,  q^{1/3} (1+q+2q^2+2q^4+q^5+2q^6+\ldots) \,,\\
\label{eq:Y1-Y2-q-exp}Y_2(\tau)&=&-\frac{\eta^{3}(3\tau)}{\eta(\tau)}-\frac{1}{3}\frac{\eta^{3}(\tau / 3)}{\eta(\tau)}=-1-6q-6q^3-6q^4-12q^7-6q^9+\ldots\,.
\end{eqnarray}

\begin{table}[hp!]
\centering
\begin{tabular}{|c|c|}\hline\hline
Modular weight $k$ & Modular form multiplets $Y_{\mathbf{r}}^{(k)}(\tau)$ \\\hline

$k=1$ &  $Y^{(1)}_{\bm{\widehat{2}}}(\tau)=\left(Y_1(\tau)\,, Y_2(\tau)\right)^T$ \\ \hline

$k=2$ &  $Y^{(2)}_{\bm{3}}(\tau)=\left(Y^{(1)}_{\bm{\widehat{2}}}\otimes Y^{(1)}_{\bm{\widehat{2}}}\right)_{\bm{3}}=\left(Y^2_2(\tau),\quad \sqrt{2}\,Y_1(\tau)Y_2(\tau),\quad -Y^2_1(\tau)\right)^T$ \\ \hline

$k=3$ & $\begin{aligned}
&Y^{(3)}_{\bm{\widehat{2}}}=\left(Y^{(1)}_{\bm{\widehat{2}}}\otimes Y^{(2)}_{\bm{3}}\right)_{\bm{\widehat{2}}}= \left(\sqrt{3}Y_1Y^2_2 ,\quad -\sqrt{\frac{2}{3}}Y^3_1-\sqrt{\frac{1}{3}}Y^3_2 \right)^T,\,\\
&Y^{(3)}_{\bm{\widehat{2}'}}=\left(Y^{(1)}_{\bm{\widehat{2}}}\otimes Y^{(2)}_{\bm{3}}\right)_{\bm{\widehat{2}'}}=\left(0,\quad 0 \right)^T,\,\\
&Y^{(3)}_{\bm{\widehat{2}''}}=\left(Y^{(1)}_{\bm{\widehat{2}}}\otimes Y^{(2)}_{\bm{3}}\right)_{\bm{\widehat{2}''}}=\left(-\sqrt{\frac{1}{3}}Y^3_1+\sqrt{\frac{2}{3}}Y^3_2,\quad \sqrt{3}Y_2Y^2_1 \right)^T \,.
\end{aligned}$ \\ \hline

$k=4$ ~&~ $\begin{aligned}
&Y^{(4)}_{\bm{3}}=\left(Y^{(1)}_{\bm{\widehat{2}}}\otimes Y^{(3)}_{\bm{\widehat{2}}}\right)_{\bm{3}} = \frac{1}{\sqrt{3}}\left(-\sqrt{2}Y^3_1Y_2-Y^4_2,~ -Y^4_1+\sqrt{2}Y_1Y^3_2, ~ -3Y^2_1Y^2_2 \right)^T,\,\\
&Y^{(4)}_{\bm{1'}}=\left(Y^{(1)}_{\bm{\widehat{2}}}\otimes Y^{(3)}_{\bm{\widehat{2}}}\right)_{\bm{1'}}=\frac{1}{\sqrt{3}}\left(-Y^4_1-2\sqrt{2}Y_1Y^3_2\right) \,, \,\\
&Y^{(4)}_{\bm{1}}=\left(Y^{(1)}_{\bm{\widehat{2}}}\otimes Y^{(3)}_{\bm{\widehat{2}''}}\right)_{\bm{1}}=\frac{1}{\sqrt{3}}\left( 2\sqrt{2}Y^3_1Y_2-Y^4_2\right) \,.
\end{aligned} $   \\ \hline

$k=5$ ~&~ $
\begin{aligned}
& Y^{(5)}_{\bm{\widehat{2}}}=\left(Y^{(1)}_{\bm{\widehat{2}}}\otimes Y^{(4)}_{\bm{3}}\right)_{\bm{\widehat{2}}}= \frac{1}{3}\left[-2\sqrt{2}Y_{1}^{3}Y_{2}+Y_{2}^{4}\right]\left(Y_{1},\quad Y_{2}\right)^T,\,\\
& Y^{(5)}_{\bm{\widehat{2}'}}=\left(Y^{(1)}_{\bm{\widehat{2}}}\otimes Y^{(4)}_{\bm{3}}\right)_{\bm{\widehat{2}'}}=\frac{1}{3}\left[-Y_{1}^{4}-2\sqrt{2}Y_{1}Y_{2}^{3}\right]\left(Y_{1},\quad Y_{2}\right)^T,\,\\
&Y^{(5)}_{\bm{\widehat{2}''}}=\left(Y^{(1)}_{\bm{\widehat{2}}}\otimes Y^{(4)}_{\bm{3}}\right)_{\bm{\widehat{2}''}}=\frac{1}{3}\left(-5Y^3_1Y^2_2-\sqrt{2}Y^5_2,~ -\sqrt{2}Y^5_1+5Y^2_1Y^3_2 \right)^T \,.
\end{aligned} $  \\  \hline

$k=6$  &  $\begin{aligned}
& Y^{(6)}_{\bm{3}I}= \left(Y^{(1)}_{\bm{\widehat{2}}}\otimes Y^{(5)}_{\bm{\widehat{2}}}\right)_{\bm{3}} = \frac{1}{3}\left[-2\sqrt{2}Y_{1}^{3}Y_{2}+Y_{2}^{4}\right] \left( Y_{2}^{2},\quad \sqrt{2}Y_{1}Y_{2},\quad -Y_{1}^{2}\right)^{T},\,\\
&Y^{(6)}_{\bm{3}II}=\left(Y^{(1)}_{\bm{\widehat{2}}}\otimes Y^{(5)}_{\bm{\widehat{2}'}}\right)_{\bm{3}}=\frac{1}{3}\left[-Y_{1}^{4}-2\sqrt{2}Y_{1}Y_{2}^{3}\right] \left( -Y_{1}^{2},\quad Y_{2}^{2},\quad \sqrt{2}Y_{1}Y_{2}\right)^{T},\,\\
&Y^{(6)}_{\bm{1}}=\left(Y^{(1)}_{\bm{\widehat{2}}}\otimes Y^{(5)}_{\bm{\widehat{2}''}}\right)_{\bm{1}}=\frac{1}{3}\left( Y^6_2-Y^6_1+5\sqrt{2}Y^3_1Y^3_2\right)
\end{aligned}$  \\  \hline \hline

\end{tabular}
\caption{\label{tab:modular-forms-level-3}The summary of modular forms of level 3 with weight up to $6$.}
\end{table}

The higher weight modular forms of level 3 can be obtained from the tensor products of $Y^{(1)}_{\bm{\widehat{2}}}(\tau)$, and they can be expressed as polynomials of $Y_1$ and $Y_2$, as shown in table~\ref{tab:modular-forms-level-3}. Notice that the modular forms of odd weights transform as two-dimensional representations $\bm{\widehat{2}}$, $\bm{\widehat{2}}'$ and $\bm{\widehat{2}}''$ of $T'$, and the modular forms of even weights are arranged into the representations $\bm{1}$, $\bm{1}'$, $\bm{1}''$ and $\bm{3}$. For the two linearly independent triplet modular multiplets $Y^{(6)}_{\bm{3}I}$ and $Y^{(6)}_{\bm{3}II}$, the expressions of the $q$-expansion are given by
\begin{eqnarray}
\nonumber Y^{(6)}_{\bm{3}I}&=&
\begin{pmatrix}
\frac{1}{3}+84 q+1692 q^2 +13764 q^3+59628 q^4+176184 q^5+\dots \,\\
\,\\[-0.1in]
-2q^{1/3}(1 + 247q + 3848q^2 + 23778q^3 + 86174q^4 + 248911q^5 +\dots) \,\\
-6q^{2/3}(1 + 242q + 2645q^2 + 12244q^3 + 42728q^4 + 109446q^5 +\dots)
\end{pmatrix}\,,\\
Y^{(6)}_{\bm{3}II} &=&
\begin{pmatrix}
-72 q+432 q^2 -648  q^3-288 q^4-432 q^5+\dots \,\\
\,\\[-0.1in]
4q^{1/3}(1 + 4q - 40q^2  -36q^3 + 638q^4 -1136q^5 +\dots) \,\\
-24q^{2/3}(1 -q -28q^2 + 94q^3 -40q^4 -147q^5 +\dots)
\end{pmatrix}
\label{eq:q-expansion-of-Y6}\,.
\end{eqnarray}
In the present work, we would like to choose another set of weight 6 modular forms $Y^{(6)}_{\bm{3}A}$ and $Y^{(6)}_{\bm{3}B}$ which are linear combination of
$Y^{(6)}_{\bm{3}I}$ and $Y^{(6)}_{\bm{3}II}$ as follow,
\begin{eqnarray}
\nonumber Y^{(6)}_{\bm{3}A}(\tau)&=&\frac{\sqrt{3}}{2}\,Y^{(6)}_{\bm{3}I}+ \frac{1}{2}\,Y^{(6)}_{\bm{3}II}\,, ~~~ Y^{(6)}_{\bm{3}B}(\tau)=-\frac{1}{2}\,Y^{(6)}_{\bm{3}I}+\frac{\sqrt{3}}{2}Y^{(6)}_{\bm{3}II}\,.
\end{eqnarray}
One sees that both $Y^{(6)}_{\bm{3}A}(\tau)$ and $Y^{(6)}_{\bm{3}B}(\tau)$ are along the direction $(1, 0, 0)^T$ in the limit of large $\text{Im}(\tau)$. Note that the normalization of modular form multiplets can not be fixed in the bottom-up approach, it can be multiplied by any number in principle.

\section{\label{app:patterns-masses}Possible patterns of fermion masses and rotation angles}

In this Appendix, we shall present the hierarchical patterns of the fermion masses $m_{\psi_1}$, $m_{\psi_2}$, $m_{\psi_3}$ and the rotation angles $\theta^{\psi}_{12}$, $\theta^{\psi}_{13}$, $\theta^{\psi}_{23}$ as well as the permutation $P_{\psi}$ which can be obtained from the finite modular groups $\Gamma'_3\cong T'$, $\Gamma'_4\cong S'_4$ and $\Gamma'_5\cong A'_5$, see tables~\ref{tab:A4p-patterns}, \ref{tab:S4p-patterns}, and \ref{tab:A5p-patterns}. The three generations of matter fields $\psi$ and $\psi^{c}$ are assigned to three-dimensional (possible reducible) representations $\bm{r}_{\psi}$ and $\bm{r}_{\psi^c}$ of $\Gamma'_N$ respectively. We require at least one of $\bm{r}_{\psi}$ and $\bm{r}_{\psi^c}$ to be an irreducible triplet of $\Gamma'_N$, otherwise there are too many cases to be tabulated.
Notice that the general results in these tables may break down in certain models when accidental cancellation happens.

In order to accommodate the large lepton mixing angles, the electroweak lepton doublets are usually assigned to an irreducible triplet of the finite modular group $\Gamma'_{N}$. The hierarchical patterns of neutrino masses $m_{\nu_i}$ and rotation angles $\theta^{\nu}_{ij}$ are listed in table~\ref{tab:neutrino-patterns} under the assumption that the light neutrino masses are described by the Weinberg operator.

\begin{center}
\setlength\LTleft{0.1in}
\setlength\LTright{0pt}
\renewcommand{\arraystretch}{1.5}
\footnotesize
\begin{longtable}{|c|c|c|c|c|}
\caption{\label{tab:A4p-patterns}
The order of magnitudes of fermion masses $m_{\psi_i}$ and rotation angles $\theta^{\psi}_{ij}$ as well as the permutation matrix $P_{\psi}$, where the three generations of matter fields  $\psi\equiv(\psi_1, \psi_2,\psi_3)^T$ and $\psi^c\equiv(\psi^c_1, \psi^c_2,\psi^c_3)^T$ are assigned to three-dimensional (possibly reducible) representations of the finite modular group $\Gamma_{3}^{\prime}\cong T^{\prime}$. For example, $\bm{r}_{\psi^{c}}=\bm{1}^{\prime\prime}\oplus\bm{1}^{\prime}\oplus\bm{1}$ means
$\bm{r}_{\psi_1^{c}}=\bm{1}^{\prime\prime}$, $\bm{r}_{\psi_2^{c}}=\bm{1}^{\prime}$,
$\bm{r}_{\psi_3^{c}}=\bm{1}$.} \\ \hline
$\bm{r}_{\psi}$ $\otimes$ $\bm{r}_{\psi^{c}}$ &power of $\tilde{\phi}$ &$P_{\psi}$
&$\left(m_{\psi_1}, m_{\psi_2}, m_{\psi_3} \right)$
&$(\theta^{\psi}_{12},\theta^{\psi}_{23},\theta^{\psi}_{13})
$\,\\ \hline
\endfirsthead

\multicolumn{5}{c}
{{\bfseries \tablename\ \thetable{} -- continued from previous page}} \\ \hline
$\bm{r}_{\psi}$ $\otimes$ $\bm{r}_{\psi^{c}}$ &power of $\tilde{\phi}$ &$P_{\psi}$
&$(m_{\psi_1}, m_{\psi_2}, m_{\psi_3})$ &$(\theta^{\psi}_{12},\theta^{\psi}_{23},\theta^{\psi}_{13})$\,\\ \hline
\endhead

\hline
\multicolumn{5}{|r|}{{continues on next page}}\\  \hline
\endfoot

\hline\hline
\endlastfoot

$\bm{3}$ $\otimes$ $\bm{3}$
& $I=J=K$
&$P_{132}$
&$(\tilde{\phi}^{I},\tilde{\phi}^{I},\tilde{\phi}^{I})$
&$(|q|^{\frac{1}{3}},|q|^{\frac{1}{3}},|q|^{\frac{1}{3}})$               \,\\ \hline
\multirow{6}{*}{$\bm{3}$ $\otimes$ $
(\bm{1}^{\prime\prime}
\oplus
\bm{1}^{\prime}
\oplus
\bm{1})$}
&$I\geq J\geq K$
&$P_{231}$
& $\left(\tilde{\phi}^{I},\tilde{\phi}^{J},
\tilde{\phi}^{K}\right)$
&  $(|q|^{\frac{1}{3}}
,|q|^{\frac{1}{3}}
,|q|^{\frac{1}{3}}\tilde{\phi}^{2(I-K)}+|q|^{\frac{2}{3}})$         \,\\*
       &
$J\geq K\geq I$
&$P_{312}$
& $(\tilde{\phi}^{J},\tilde{\phi}^{K},\tilde{\phi}^{I})$
&$(|q|^{\frac{1}{3}},|q|^{\frac{1}{3}}
,|q|^{\frac{1}{3}}\tilde{\phi}^{2(J-I)}+|q|^{\frac{2}{3}})
$         \,\\*
       &
$K\geq I\geq J$
&$P_{123}$
& $(\tilde{\phi}^{K},\tilde{\phi}^{I},\tilde{\phi}^{J})$               &$(|q|^{\frac{1}{3}},|q|^{\frac{1}{3}}
,|q|^{\frac{1}{3}}\tilde{\phi}^{2(K-J)}+|q|^{\frac{2}{3}})$         \,\\*
       &
$J\geq I\geq K$
&$P_{321}$
& $(\tilde{\phi}^{J},\tilde{\phi}^{I},\tilde{\phi}^{K})$               &$(|q|^{\frac{2}{3}}
,|q|^{\frac{1}{3}}\tilde{\phi}^{2(I-K)}+|q|^{\frac{2}{3}}
,|q|^{\frac{1}{3}})$         \,\\*
       &
$I\geq K\geq J$
&$P_{213}$
& $(\tilde{\phi}^{I},\tilde{\phi}^{K},\tilde{\phi}^{J})$               &$(|q|^{\frac{2}{3}}
,|q|^{\frac{1}{3}}\tilde{\phi}^{2(K-J)}+|q|^{\frac{2}{3}}
,|q|^{\frac{1}{3}})$         \,\\*
       &
$K\geq J\geq I$
&$P_{132}$
& $(\tilde{\phi}^{K},\tilde{\phi}^{J},\tilde{\phi}^{I})$               &$(|q|^{\frac{2}{3}}
,|q|^{\frac{1}{3}}\tilde{\phi}^{2(J-I)}+|q|^{\frac{2}{3}}
,|q|^{\frac{1}{3}})$         \,\\ \hline
\multirow{8}{*}{$\bm{3}$ $\otimes$ $
(\bm{1}^{\prime}
\oplus
\bm{1}
\oplus
\bm{1})$}
&$
I\geq J\geq K,
|q|^{\frac{1}{3}}\tilde{\phi}^{J}
\geq\tilde{\phi}^{I}
$
&$P_{231}$
& $(|q|^{\frac{1}{3}}\tilde{\phi}^{I},
|q|^{\frac{1}{3}}\tilde{\phi}^{J},\tilde{\phi}^{K})$               &$(|q|^{\frac{1}{3}},|q|^{\frac{1}{3}},|q|^{\frac{2}{3}})$         \,\\
\hline
\end{longtable}
\end{center}

\setlength\LTleft{-0.38in}
\setlength\LTright{0pt}
\setlength\LTleft{-0.6in}
\setlength\LTright{-2pt}
\renewcommand{\arraystretch}{1.5}
\begin{center}
\footnotesize
\begin{longtable}{|c|c|c|c|c|}
\caption{\label{tab:S4p-patterns}
The order of magnitudes of fermion masses $m_{\psi_i}$ and rotation angles $\theta^{\psi}_{ij}$ as well as the permutation matrix $P_{\psi}$, where the three generations of matter fields  $\psi\equiv(\psi_1, \psi_2,\psi_3)^T$ and $\psi^c\equiv(\psi^c_1, \psi^c_2,\psi^c_3)^T$ are assigned to three-dimensional (possibly reducible) representations of the finite modular group $\Gamma_{4}^{\prime}\cong S^{\prime}_4$. Here we adopt the basis of Ref.~\cite{Ding:2023htn} for the representation matrices of $S^{\prime}_4$. }\,\\\hline
$\bm{r}_{\psi}$$\otimes$ $\bm{r}_{\psi^{c}}$ &power of $\tilde{\phi}$ &$P_{\psi}$
&$(m_{\psi_1}, m_{\psi_2}, m_{\psi_3})$ &$(\theta^{\psi}_{12},\theta^{\psi}_{23},\theta^{\psi}_{13})$\,\\ \hline
\endfirsthead

\multicolumn{5}{c}
{{\bfseries \tablename\ \thetable{} -- continued from previous page}} \\ \hline
$\bm{r}_{\psi}$ $\otimes$ $\bm{r}_{\psi^{c}}$ &power of $\tilde{\phi}$ &$P_{\psi}$
& $(m_{\psi_1}, m_{\psi_2}, m_{\psi_3})$ & $(\theta^{\psi}_{12},\theta^{\psi}_{23},\theta^{\psi}_{13})$\, \\ \hline
\endhead

\hline
\multicolumn{5}{|r|}{{continues on next page}}\\  \hline
\endfoot

\hline
\endlastfoot
$\bm{3}^{\prime}$    $\otimes$ $\bm{3}^{\prime}$
& $I=J=K$
&$P_{213}$
&$(\tilde{\phi}^{J},\tilde{\phi}^{J},\tilde{\phi}^{J})$
&$(|q|^{\frac{1}{4}},|q|^{\frac{1}{4}},|q|^{\frac{1}{2}})$                \,\\ \hline
$\bm{3}^{\prime}$    $\otimes$ $\bm{3}$
& $I=J=K$
&$P_{123}$
&$(|q|^{\frac{1}{2}}\tilde{\phi}^{J},
\tilde{\phi}^{J},\tilde{\phi}^{J})$
&$(|q|^{\frac{1}{4}},|q|^{\frac{1}{2}},|q|^{\frac{3}{4}})$                \,\\ \hline
$\bm{3}^{\prime}$    $\otimes$ $\widehat{\bm{3}}^{\prime}$
& $I=J=K$
&$P_{321}$
&$(|q|^{\frac{3}{4}}\tilde{\phi}^{J},
\tilde{\phi}^{J},\tilde{\phi}^{J})$
&$(|q|^{\frac{1}{2}},|q|^{\frac{1}{4}},|q|^{\frac{1}{4}})$                    \,\\ \hline
$\bm{3}^{\prime}$    $\otimes$ $\widehat{\bm{3}}$
& $I=J=K$
&$P_{213}$
&$(|q|^{\frac{1}{4}}\tilde{\phi}^{J},
\tilde{\phi}^{J},\tilde{\phi}^{J})$
&$(|q|^{\frac{3}{4}},|q|^{\frac{1}{4}},|q|^{\frac{1}{2}})$                  \,\\ \hline
$\bm{3}$    $\otimes$ $\bm{3}$
& $I=J=K$
&$P_{132}$
&$(\tilde{\phi}^{J},
\tilde{\phi}^{J},\tilde{\phi}^{J})$
&$(|q|^{\frac{1}{4}},|q|^{\frac{1}{2}},|q|^{\frac{1}{4}})$                 \,\\ \hline
$\bm{3}$    $\otimes$ $\widehat{\bm{3}}$
& $I=J=K$
&$P_{312}$
&$(|q|^{\frac{3}{4}}\tilde{\phi}^{J},
\tilde{\phi}^{J},\tilde{\phi}^{J})$
&$(|q|^{\frac{1}{4}},|q|^{\frac{1}{4}},|q|^{\frac{1}{2}})$     \,\\ \hline
$\widehat{\bm{3}}^{\prime}$    $\otimes$ $\bm{3}$
& $I=J=K$
&$P_{231}$
&$(|q|^{\frac{1}{4}}\tilde{\phi}^{J},
\tilde{\phi}^{J},\tilde{\phi}^{J})$
&$(|q|^{\frac{1}{2}},|q|^{\frac{1}{4}},|q|^{\frac{3}{4}})$    \,\\ \hline
$\widehat{\bm{3}}^{\prime}$    $\otimes$ $\widehat{\bm{3}}^{\prime}$
& $I=J=K$
&$P_{132}$
&$(|q|^{\frac{1}{2}}\tilde{\phi}^{J},
\tilde{\phi}^{J},\tilde{\phi}^{J})$
&$(|q|^{\frac{3}{4}},|q|^{\frac{1}{2}},|q|^{\frac{1}{4}})$                  \,\\
\hline
\end{longtable}
\end{center}

\begin{center}
\footnotesize

\setlength\LTleft{0in}
\setlength\LTright{0pt}
\renewcommand{\arraystretch}{1.5}
\begin{longtable}{|c|c|c|c|c|}
\caption{\label{tab:A5p-patterns}
The order of magnitudes of fermion masses $m_{\psi_i}$ and rotation angles $\theta^{\psi}_{ij}$ as well as the permutation matrix $P_{\psi}$, where the three generations of matter fields  $\psi\equiv(\psi_1, \psi_2,\psi_3)^T$ and $\psi^c\equiv(\psi^c_1, \psi^c_2,\psi^c_3)^T$ are assigned to three-dimensional (possibly reducible) representations of the finite modular group $\Gamma_{5}^{\prime}\cong A^{\prime}_5$.  Here we adopt the basis of Ref.~\cite{Ding:2023htn} for the representation matrices of $A^{\prime}_5$. }\,\\ \hline
$\bm{r}_{\psi}$ $\otimes$ $\bm{r}_{\psi^{c}}$ & power of $\tilde{\phi}$ & $P_{\psi}$
& $(m_{\psi_1}, m_{\psi_2}, m_{\psi_3})$ & $(\theta^{\psi}_{12},\theta^{\psi}_{23},\theta^{\psi}_{13})$ \,\\ \hline
\endfirsthead

\multicolumn{5}{c}{{\bfseries \tablename\ \thetable{} -- continued from previous page}} \\ \hline
$\bm{r}_{\psi}$ $\otimes$ $\bm{r}_{\psi^{C}}$ & power of $\tilde{\phi}$ & $P_{\psi}$
&$(m_{\psi_1}, m_{\psi_2}, m_{\psi_3})$
&$(\theta^{\psi}_{12},\theta^{\psi}_{23},\theta^{\psi}_{13})
$ \,\\ \hline
\endhead

\hline
\multicolumn{5}{|r|}{{continues on next page}}\\  \hline
\endfoot

\hline
\endlastfoot
$\bm{3}$    $\otimes$ $\bm{3}$
& $I=J=K$
&$P_{132}$
&$(\tilde{\phi}^{I},\tilde{\phi}^{I},\tilde{\phi}^{I})$
&$(|q|^{\frac{1}{5}},|q|^{\frac{2}{5}},|q|^{\frac{1}{5}})$                \,\\ \hline
$\bm{3}$    $\otimes$ $\bm{3}^{\prime}$
& $I=J=K$
&$P_{321}$
&$(|q|^{\frac{4}{5}}\tilde{\phi}^{I},
|q|^{\frac{1}{5}}\tilde{\phi}^{I},\tilde{\phi}^{I})$
&$(|q|^{\frac{2}{5}},|q|^{\frac{3}{5}},|q|^{\frac{1}{5}})$    \,\\ \hline
$\bm{3}^{\prime}$    $\otimes$ $\bm{3}^{\prime}$
& $I=J=K$
&$P_{213}$
&$(\tilde{\phi}^{I},\tilde{\phi}^{I},\tilde{\phi}^{I})$
&$(|q|^{\frac{2}{5}},|q|^{\frac{2}{5}},|q|^{\frac{1}{5}})$    \,\\ \hline
$\bm{3}$    $\otimes$ $(\bm{1}\oplus{\bm{1}}\oplus{\bm{1}})$
&$I\geq J\geq K$
&$P_{231}$
& $(|q|^{\frac{4}{5}}\tilde{\phi}^{I},
|q|^{\frac{1}{5}}\tilde{\phi}^{J},\tilde{\phi}^{K})$
&$(|q|^{\frac{3}{5}},|q|^{\frac{1}{5}},|q|^{\frac{4}{5}})$     \,\\ \hline
$\bm{3}^{\prime}$    $\otimes$ $(\bm{1}\oplus{\bm{1}}\oplus{\bm{1}})$
&$I\geq J\geq K$
&$P_{231}$
& $(|q|^{\frac{3}{5}}\tilde{\phi}^{I},
|q|^{\frac{2}{5}}\tilde{\phi}^{J},\tilde{\phi}^{K})$
&$(|q|^{\frac{1}{5}},|q|^{\frac{2}{5}},|q|^{\frac{3}{5}})$     \,\\\hline

\multirow{4}{*}{$\bm{3}$   $\otimes$ $(\widehat{\bm{2}}\oplus{\bm{1}})$}
&$I=J\geq K$
&$P_{321}$
& $(|q|^{\frac{4}{5}}\tilde{\phi}^{J},
|q|^{\frac{1}{5}}\tilde{\phi}^{J},\tilde{\phi}^{K})$
&$(|q|^{\frac{2}{5}}
,|q|^{\frac{4}{5}}+|q|^{\frac{3}{5}}\tilde{\phi}^{2(J-K)}
,|q|^{\frac{1}{5}})$              \,\\*
       &
$K\geq I=J,
~|q|^{\frac{3}{5}}\tilde{\phi}^{J}\geq\tilde{\phi}^{K}$
&$P_{312}$
& $(|q|^{\frac{1}{5}}\tilde{\phi}^{K},
|q|^{\frac{3}{5}}\tilde{\phi}^{J},|q|^{\frac{1}{5}}
\tilde{\phi}^{J})$
&$(|q|^{\frac{1}{5}}
,|q|^{\frac{1}{5}}
,|q|^{\frac{2}{5}})$            \,\\
\hline
\end{longtable}
\end{center}

\begin{table}[htb!]
\centering
\begin{tabular}{|c|c|c|c|c|c|}\hline\hline
Modular group
&$\bm{r}_{\nu}$
&power of $\tilde{\phi}$
&$P_{\nu}$
&$(m_{\nu_1}, m_{\nu_2}, m_{\nu_3})$
&$(\theta^{\nu}_{12}, \theta^{\nu}_{23},\theta^{\nu}_{13})$\,\\ \hline

$T'$  & $\bm{3}$ & $I=J=K$ &$O_{23}$
&$(\tilde{\phi}^{I},\tilde{\phi}^{I},\tilde{\phi}^{I})$
&$(|q|^{\frac{1}{3}},|q|^{\frac{1}{3}},|q|^{\frac{1}{3}})$     \,\\ \hline
$S_{4}^{\prime}$   &$\bm{3}^{(\prime)}$
& $I=J=K$
&$O_{23}$ & $(\tilde{\phi}^{I},\tilde{\phi}^{I},\tilde{\phi}^{I})$
&$(|q|^{\frac{1}{4}},|q|^{\frac{1}{2}},|q|^{\frac{1}{4}})$  \,\\ \hline
$S_{4}^{\prime}$   &$\widehat{\bm{3}}^{(\prime)}$
& $I=J=K$
&$P_{123}$
&$(|q|^{\frac{1}{2}}\tilde{\phi}^{I},
\tilde{\phi}^{I},\tilde{\phi}^{I})$
&$(|q|^{\frac{1}{4}},|q|^{\frac{1}{2}},|q|^{\frac{3}{4}})$ \,\\ \hline
$A_{5}^{\prime}$   &$\bm{3}$
& $I=J=K$
&$O_{23}$
&$(\tilde{\phi}^{I},\tilde{\phi}^{I},\tilde{\phi}^{I})$
&$(|q|^{\frac{1}{5}},|q|^{\frac{2}{5}},|q|^{\frac{1}{5}})$     \,\\ \hline
$A_{5}^{\prime}$   &$\bm{3}^{\prime}$
& $I=J=K$
&$O_{23}$
&$(\tilde{\phi}^{I},\tilde{\phi}^{I},\tilde{\phi}^{I})$
&$(|q|^{\frac{2}{5}},|q|^{\frac{1}{5}},|q|^{\frac{2}{5}})$ \,\\  \hline\hline
\end{tabular}
\caption{\label{tab:neutrino-patterns}The order of magnitudes of neutrino masses $m_{\nu_i}$ and rotation angles $\theta^{\nu}_{ij}$ as well as the permutation matrix $P_{\nu}$, where the light neutrino masses are generated by the Weinberg operator and the electroweak lepton doublets are assigned to an irreducible triplet $\bm{r}_{\nu}$ of the finite modular group $\Gamma'_{N}$. }
\end{table}

\section{\label{sec:Yukawa-CP-fixed-points}Yukawa couplings in the vicinity of residual CP symmetric points}

In this Appendix, we shall study the
transformation of the Yukawa couplings under the gCP symmetry and show that the Yukawa couplings are subject to the constraint from the residual CP symmetry in the vicinity of CP conserved points. The modular symmetry can be consistently combined with the gCP symmetry. By considering the consistency condition chain between the modular symmetry and CP symmetry, one can show that the CP transformation of the complex modulus $\tau$ is $\tau\xrightarrow{\mathcal{CP}}-\bar{\tau}$ up to modular transformations~\cite{Dent:2001cc,Dent:2001mn,Baur:2019kwi,Baur:2019iai,Novichkov:2019sqv}. This CP symmetry corresponds to an outer automorphism of the modular group: $u(\gamma)=\mathcal{CP}\circ\gamma\circ\mathcal{CP}^{-1}=\begin{pmatrix}
a  & -b\\
-c & d
\end{pmatrix}$, which implies $u(S)=S^{-1}$ and $u(T)=T^{-1}$~\cite{Novichkov:2019sqv}.
The CP transformation of the modulus and the matter fields are given by
\begin{eqnarray}
\tau\stackrel{\mathcal{CP}}{\mapsto}-\bar{\tau},~\quad
~L\stackrel{\mathcal{CP}}{\mapsto}X_{L}\overline{L},~\quad~
E^c\stackrel{\mathcal{CP}}{\mapsto}X_{E^c}\overline{E^c}\,~\quad~H_u\stackrel{\mathcal{CP}}{\mapsto}X_{u}\overline{H_u},~\quad~
H_d\stackrel{\mathcal{CP}}{\mapsto}X_{d}\overline{H_d}\,,
\end{eqnarray}
where $X_{L}$, $X_{E^c}$, $X_{u}$ and $X_{d}$ are the CP transformation matrices of the fields $L$, $E^c$, $H_u$ and $H_d$ respectively and the flavor indices are dropped, their explicit forms depend on the modular transformation of these fields. In the basis where both $S$ and $T$ are represented by symmetric and unitary matrices in any irreducible representations of $\Gamma'_N$ or $\Gamma_N$, the CP transformation matrix coincides with the unity matrix and consequently the action of CP symmetry on matter multiplet $\Phi_I$ reduces to the canonical one: $\Phi_I\xrightarrow{\mathcal{CP}}\bar{\Phi}_I$~\cite{Novichkov:2019sqv}. In such basis, the CP invariance restricts all couplings in the superpotential to be real, if the Clebsch-Gordan coefficients are real and the modular forms are properly normalized~\cite{Novichkov:2019sqv}.

It is straightforward to derive that invariance of $\mathcal{W}_m$ under CP transformation requires:
\begin{eqnarray}
\mathcal{Y}^e(-\tau^*)&= X_d^{*}~X_{E^c}^*~\mathcal{Y}^{e*}(\tau)~ X_L^\dagger\,,~~~\mathcal{Y}^{\nu}(-\tau^*)=X_u^{*2}~X_{L}^*~\mathcal{Y}^{\nu*}(\tau)~ X_L^\dagger~~~\,.
\label{eq:me-mnu-CP}
\end{eqnarray}
Combining the modular transformation Eq.~\eqref{eq:me-mnu-gamma} and CP transformation Eq.~\eqref{eq:me-mnu-CP}, we obtain
\begin{eqnarray}
\nonumber\mathcal{Y}^{e}_{ij}\Bigl(\gamma(-\tau^{\ast})\Bigr)&=&(-c\tau^{\ast}+d)^{k_{E^{c}_i}+k_{L_j}+k_{d}}\rho^{\ast}_{d}(\gamma)X^{\ast}_{d}
\Bigl[\rho_{E^{c}}^{\ast}(\gamma)X^{*}_{E^{c}}\mathcal{Y}^{e*}(\tau)X_{L}^{\dag}
\rho_{L}^{\dag}(\gamma)\Bigr]_{ij}\,,~~~\\
\mathcal{Y}^{\nu}_{ij}\Bigl(\gamma(-\tau^{\ast})\Bigr)&=&(-c\tau^{\ast}+d)^{k_{L_{i}}+k_{L_{j}}+2k_{u}}\rho^{*2}_{u}(\gamma)X^{\ast2}_{u}
\Bigl[\rho_{L}^{\ast}(\gamma)X_{L}^{\ast}\mathcal{Y}^{\nu\ast}(\tau)
X_{L}^{\dag}\rho_{L}^{\dag}(\gamma)\Bigr]_{ij}\,. \label{eq:me-mnu-CP-modsym}
\end{eqnarray}

\begin{figure}[t!]
\centering
\includegraphics[width=0.50\textwidth]{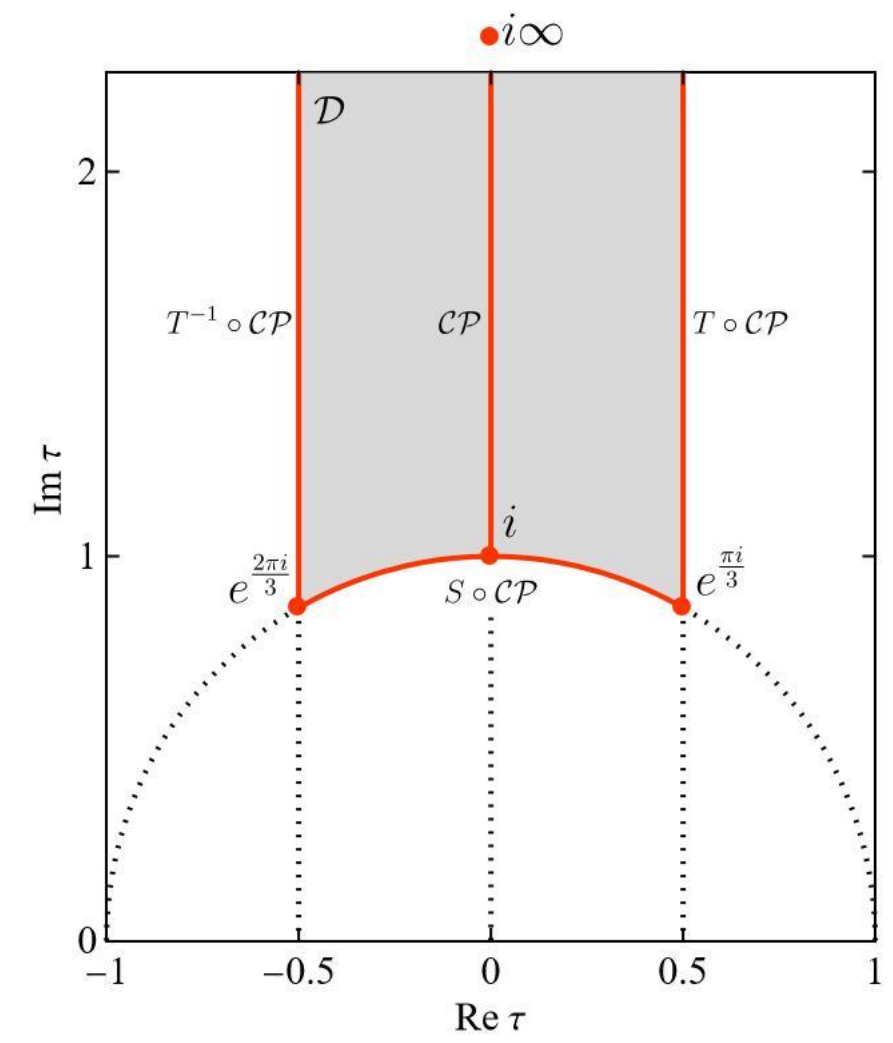}
\caption{\label{fig:gCP-fixed-points}The fixed points and CP conserving boundary in the fundamental domain $\mathcal{D}$ of the modulus field $\tau$ is displayed as the red points and red lines and the corresponding residual CP symmetries are indicated. The intersection fixed points $\tau=e^{\pi i/3},\, i, e^{2\pi i/3}$ enjoy two different residual CP symmetries so that a residual modular symmetry can be generated. For instance, the self-dual point $\tau=i$ is invariant under both CP transfromations $Z^{\mathcal{CP}}_2$ and $Z^{S\circ\mathcal{CP}}_2$, consequently it is also invariant under the modular generator $S$.  }
\end{figure}

The combination of any modular transformation $\gamma$ with CP transformation gives another allowed CP transformation,
\begin{equation}
\label{eq:gCP-transf}\tau\xrightarrow{\gamma}\frac{a\tau+b}{c\tau+d}\xrightarrow{\mathcal{CP}}\frac{-a\tau^{*}+b}{-c\tau^{*}+d}\equiv (\gamma\circ\mathcal{CP})\tau\,.
\end{equation}
There are some point $\tau_0$ in the fundamental domain which is invariant under the action of the general CP transformation $\gamma_0\circ\mathcal{CP}$~\cite{Novichkov:2019sqv}, i.e.,
\begin{equation}
\gamma_{0}\circ\mathcal{CP}(\tau_0)=\tau_0\,.
\end{equation}
The left vertical boundary, the right vertical boundary, the unit arc boundary of $\mathcal{D}$ and the imaginary axis are invariant under the action of $T^{-1}\circ\mathcal{CP}$, $T\circ\mathcal{CP}$, $S\circ\mathcal{CP}$ and $\mathcal{CP}$ corresponding to $\gamma_0=T^{-1}, T, S, 1$ respectively~\cite{Novichkov:2019sqv,Ding:2023htn}, as shown in figure~\ref{fig:gCP-fixed-points}. The modular symmetry fixed points are the intersection of the CP conserved lines. The CP invariance is restored at these residual CP fixed points so that all the CP violation phases are conserved (equal to 0 or $\pi$)~\cite{Novichkov:2019sqv}. Hence departure of $\langle\tau\rangle$ from residual CP symmetry fixed points is required to accommodate a non-trivial CP violation. Notice that even a small deviation can lead to large CP violation due to strong dependence of the observables on $\langle\tau\rangle$~\cite{Novichkov:2019sqv}.

By taking $\gamma=\gamma_0$ and $\tau=\tau_0$ in Eq.~\eqref{eq:me-mnu-CP-modsym}, we see that the hermitian combination of the neutrino and charged lepton mass matrices $M^{\dagger}_{\nu}(\tau_0)M_{\nu}(\tau_0)$ and $M^{\dagger}_{e}(\tau_0)M_{e}(\tau_0)$ are invariant under a common transformation of the left-handed charged leptons and the left-handed neutrinos $\Omega_L=(-c_0\tau^{*}_0+d_0)^{-k_{L}}\rho_{L}(\gamma_0)X_{L}$, which represents a combination of CP and modular transformations. The unitary transformation $\Omega_L$ should be symmetric otherwise the neutrino and the charged lepton mass spectrum would be constrained to be partially degenerate~\cite{Feruglio:2012cw,Chen:2014wxa,Chen:2015nha,Ding:2024ozt}. As a consequence, when the complex modulus is the residual CP fixed point with $\tau=\tau_0$, all the Dirac and Majorana CP phases would be trivial. Therefore values of modulus deviating from residual CP symmetry fixed points are required to accommodate the observed non-degenerate lepton masses and a non-trivial Dirac CP. Moreover, it was found that the minima of the modular invariant scalar potential is close to but distinct from the CP fixed points~\cite{Cvetic:1991qm,Novichkov:2022wvg,Leedom:2022zdm,Ishiguro:2020tmo,Ishiguro:2022pde,Knapp-Perez:2023nty}. Cosmological evolution can offer a mechanism for moduli trapping near the points enjoying an enhanced symmetry~\cite{Kofman:2004yc,Enomoto:2013mla,Kikuchi:2023uqo,Kobayashi:2023spx,Abe:2024tox}.

It is convenient to parameterize the small deviation of $\tau$ from the CP boundary by the parameter $\epsilon$ defined as~\cite{Feruglio:2021dte,Novichkov:2021evw,Feruglio:2022koo,Feruglio:2023mii,Ding:2024xhz},
\begin{equation}
\epsilon\equiv e^{i\alpha}\frac{\tau-\tau_{0}}
 {\tau-\tau_{0}^{\ast}}\,,
\end{equation}
where the phase $\alpha$ can be conveniently chosen to simplify the action of CP. In the following, we discuss the property of Yukawa couplings around the CP boundary.

\subsection{$\tau_{0}=iy$ with $y\geq1$ }

The imaginary axis is invariant under the CP transformation with $\gamma_0=\mathbb{1}$. We choose the phase $\alpha=0$, then the CP transformation of the departure parameter $\epsilon$ is $\epsilon\stackrel{\mathcal{CP}}{\longrightarrow}\epsilon^{*}$. Setting $\gamma=\mathbb{1}$ in Eq.~\eqref{eq:me-mnu-CP-modsym}, one finds
\begin{equation}
 \mathcal{Y}^{e}_{ij}(\epsilon^{\ast})=\mathcal{Y}^{e\ast}_{ij}(\epsilon)\,,
\end{equation}
Expanding the Yukawa couplings $\mathcal{Y}^{e}_{ij}(\epsilon)=\sum^{\infty}_{n=0}a^{ij}_{n}\epsilon^{n}$ in powers of $\epsilon$, one obtains the coefficients
\begin{equation}
a^{ij}_{n}=a^{ij\ast}_{n}\,.
\end{equation}
Therefore the CP symmetry constrains all coefficients $a^{ij}_n$ to be real, while the power index $n$ is not subject to any constraint.

\subsection{$\tau_{0}=-\frac{1}{2}+iy$ with $y\geq\frac{\sqrt{3}}{2}$ }

The left vertical boundary of the fundamental domain preserves the residual gCP symmetry $T^{-1}\circ\mathcal{CP}$ with $\gamma_0=T^{-1}$. It is convenient to take $\alpha=0$, the gCP transformation of $\epsilon$ is determined to be $\epsilon\stackrel{T^{-1}\circ\mathcal{CP}}{\longrightarrow}\epsilon^{*}$.
Setting $\gamma=T^{-1}$ in Eq.~\eqref{eq:me-mnu-CP-modsym}, one has
\begin{equation}
\label{eq:gCP-MF-left-bd}\mathcal{Y}^{e}_{ij}\left(\epsilon^{\ast}\right)=
\rho^{\ast}_{d}(T^{-1})\Bigl[\rho_{E^{c}}^{\ast}(T^{-1}) \mathcal{Y}^{e*}(\epsilon)\rho_{L}^{\dag}(T^{-1})\Bigr]_{ij}\,
\end{equation}
Considering the basis in which $T$ is represented by a diagonal matrix
$\rho(T)=\text{diag}(\zeta^{p_1}, \zeta^{p_2}, \ldots)$ with $p_i=0, 1, \ldots, N-1$, then the CP invariance imposes the following constraint on the Yukawa couplings,
\begin{equation}
\mathcal{Y}^{e}_{ij}\left(\epsilon^{\ast}\right)=\zeta^{k_{ij}}\mathcal{Y}^{e*}_{ij}(\epsilon)\,,
\end{equation}
where $\rho_{d}(T)\rho_{E^{c}_i}(T)\rho_{L_{j}}(T)=\zeta^{k_{ij}}$, $k_{ij}$ is fixed by the representation assignment of the fields $H_d$, $E^c$ and $L$, and it can take the values of $0\,,1\,,2\,,\ldots, N-1$. Expanding $\mathcal{Y}^{e}_{ij}(\epsilon)=\sum^{\infty}_{n=0}a^{ij}_{n}\epsilon^{n}$ in powers of $\epsilon$, one obtains the phase of the expansion coefficient $a^{ij}_n$ as follow,
\begin{equation}
\text{arg}\left(a^{ij}_n\right)=\frac{\pi k_{ij}}{N}~(\text{mod}~\pi)\,.
\end{equation}
The right vertical boundary of the fundamental domain is related to the left vertical boundary, consequently we shall not discuss the gCP constraint on the Yukawa couplings in the vicinity of right vertical boundary.

\subsection{$\tau_{0}=e^{i\phi}$ with $\frac{\pi}{3}\leq\phi\leq\frac{2\pi}{3}$ }

The unit arc boundary of the fundamental domain is invariant under  $S\circ\mathcal{CP}$ with $\gamma_0=S$. Choosing $\alpha=-\phi$, we find that the residual gCP acts on the departure parameter $\epsilon$ as
\begin{equation}
\epsilon\stackrel{S\circ\mathcal{CP}}{\longrightarrow}\epsilon^{\ast}\,.
\end{equation}
Setting $\gamma=S$ in Eq.~\eqref{eq:me-mnu-CP-modsym} and in terms of $\epsilon$, we obtain
\begin{equation}
\widetilde{\mathcal{Y}}^{e}_{ij}\left(\epsilon^{\ast}\right)=e^{-iK_{ij}\phi}
\rho^{\ast}_{d}(S)\Bigl[\rho_{E^{c}}^{\ast}(S)\widetilde{\mathcal{Y}}^{e*}(\epsilon)\rho_{L}^{\dag}(S)\Bigr]_{ij}\,,
~~~~\widetilde{\mathcal{Y}}^{e}_{ij}(\epsilon)=(1-e^{i\phi}\epsilon)^{-K_{ij}}\mathcal{Y}^{e}_{ij}(\epsilon)\,,
\end{equation}
with $K_{ij}\equiv k_{E^{c}_i}+k_{L_j}+k_{H_{d}}$. Expanding both sides in the $S$-diagonal basis $\rho(S)=\text{diag}(i^{l_1}, i^{l_2}, \ldots)$ with $l_i=0, 1, 2, 3$ because of $S^4=1$, one has
\begin{equation}
\widetilde{\mathcal{Y}}^{e}_{ij}\left(\epsilon^{\ast}\right)=e^{-i\left(K_{ij}\phi+k_{ij}\pi/2\right)}
\widetilde{\mathcal{Y}}^{e*}_{ij}(\epsilon)\,,
\end{equation}
where $\rho_{d}(S)\rho_{E^{c}_i}(S)\rho_{L_{j}}(S)=i^{k_{ij}}=e^{i\pi k_{ij}/2}$. Notice that invariance under the modular transformation $R=S^2$ requires $\rho_{d}(S)\rho_{E^{c}_i}(S)\rho_{L_{j}}(S)=\pm e^{i\pi K_{ij}/2}$ if the Yukawa coupling $\mathcal{Y}^{e}_{ij}(\epsilon)$ are non-zero. Expanding $\widetilde{\mathcal{Y}}^{e}_{ij}(\epsilon)=\sum^{\infty}_{n=0}a^{ij}_{n}\epsilon^{n}$, we obtain
\begin{equation}
\text{arg}\left(a^{ij}_n\right)=-\frac{K_{ij}\phi}{2}-\frac{k_{ij}\pi}{4}~(\text{mod}~\pi)\,.
\end{equation}
In short, the residual CP symmetry can constrain the phases of the expansion coefficients in the vicinity of the CP fixed points, and similar results are obtained by considering gCP transformation of modular forms.

\end{appendix}

\providecommand{\href}[2]{#2}\begingroup\raggedright\endgroup

\end{document}